\documentclass[iop, numberedappendix]{./emulateapj}
\usepackage{apjfonts}
\usepackage{epsfig,graphicx}
\usepackage{psfrag}
\usepackage[percent]{overpic}
\usepackage{color}
\usepackage{subfigure}
\usepackage{amssymb}
\usepackage{amsmath}
  \usepackage{times}
\usepackage{epstopdf}
  \usepackage{natbib}
\usepackage{url}

\newcommand{\myemail}{cabanac@cesr.fr}

\shortauthors{Cabanac et al.}
\shorttitle{Fast Timing Analysis of Cyg X-1 using SPI}

\begin{document}
\title{Fast Timing Analysis of Cygnus X-1 using SPI on board {\it INTEGRAL}}
\author{Cl\'ement~Cabanac, Jean-Pierre~Roques and Elisabeth~Jourdain}
\affil{Universit\'e de Toulouse; UPS-OMP; IRAP; Toulouse, France}
\affil{CNRS; IRAP; 9, av du Colonel Roche, BP 44346, F-31028 Toulouse Cedex 4, France}
\email{\myemail}

\begin{abstract}
For the very first time, we report the high frequency analysis of Cyg X-1 up to hard X-ray using SPI on-board {\it INTEGRAL}. After analyzing the possible contribution from the background, and using {\it INTEGRAL} archive from March 2005 to May 2008, Power Density Spectra (PDS) were obtained until 130 keV. First, we show that their overall shape is very similar to that observed at lower energies, as they are well described by sets of Lorentzians. The strength of this fast variability (up to 40 Hz) does not drop at high energy since we show that it remains at $\sim 25\%\ \rm rms$, even in the highest energy bands. Second, the hard X-ray variability patterns of Cyg X-1 are state dependent: the softer the spectrum (or the lower the hardness ratio), the lower the total fractional variability and the higher the typical frequencies observed. The strength of total variability as a function of energy and state is then investigated. By comparison with simultaneous and published {\it RXTE}/PCA data, we showed that in the hard state, it remains quite constant in the 2-130 keV energy range. In our softer state, it is also flat until 50 keV and may increase at higher energy. The implications of this behavior on the models are then discussed.
\end{abstract}


\section{Introduction}
Black Hole Binaries (BHB), either transient (such as GX 339-4) or persistent (Cyg X-1), are very variable on long time scales (i.e., from weeks to months), in terms of flux, but also spectrally speaking: since the first spectra obtained (i.e., \citealp{tananbaum1972} for Cyg X-1 and \citealp{coe1976} for A0620-00), transitions between two main states were identified. The soft state, which spectrum is peaking in the soft X-ray band (i.e., $\sim 1\ \rm keV$), is believed to be dominated by the emission of the optically thick accretion disk. For that latter reason, the soft state is also called ``Thermally Dominated'' \citep{mcclintock2006}. In the hard state, the Spectral Energy Distribution (SED) is on the contrary peaking at few tens of keV, and is thought to be emitted by an optically thin component, the so-called ``corona''. Its origin, geometry, together with the heating processes of this component still remain unclear.

The observed radio behavior of those objects is also dependent on the spectral state \citep{homan2005}: the canonical hard state is usually associated with the presence of a compact or sometimes extended steady jet (see e.g. \citealp{stirling2001} for Cyg X-1), whereas the observation of relativistic blobs motion is associated with the intermediate state. In the soft state, on the contrary, the radio emission is quenched \citep{fender1999}.  

 In order to explain such transitions, a paradigm has been set by \cite{esin1997}, where in the soft state the disc is extending down to the vicinity of the black hole, whereas it is recessed in the hard state at hundreds of gravitational radii. The inner part is in turn replaced by an advection dominated flow. Observationally, the question to know if the disc is actually receding in the hard state is still under debate (see e.g. \citealp{miller2006,reis2010}, but also \citealp{gierlinski2008,cabanac2009,tomsick2009,done2010} ).

 BHBs also exhibit strong signatures in the time domain at high frequencies. Their X-ray lightcurves are indeed characterized by erratic behavior and when analyzed in the Fourier domain, their Power Density Spectra (hereafter PDS) can show broad band aperiodic noise and/or broad incoherent peaks, the so called Quasi-Periodic Oscillations (QPO). The strength of those components, together with their typical frequency evolution, are closely dependent on the state \citep{homan2005}. For a complete review of the X-ray binaries timing characteristics, see \cite{vklis2004}.

 Since its discovery in 1964 by an Aerobee rocket, and thanks to its persistent brightness due to the wind fed accretion process, Cyg X-1 has been extensively studied. The aforementioned spectral states are indeed well identified: \cite{mcconnell2002} e.g., showed that the hard state (which is observed at a lower flux in the soft X-ray band), is well characterized by the power law component cutting off at $\sim 100\ \rm keV$, whereas the soft state (observed at higher flux in the low energy bands) is dominated by the disk emission. In addition, in both soft and hard states, a high energy tail is also observed, which is usually interpreted as the emission of non thermal particles (see e.g., \citealp{malzac2009} for a full SED modeling in presence of magnetic field, and the explanation of source state transition in terms of particle energy distribution variations).  

However, as compared to the other known X-ray binaries, Cyg X-1 is in a  number of aspect quite different. No relativistic motion has been observed in the intermediate state \citep{rushton2011} and the radio emission is not quenched in the soft state \citep{rushton2011, zdziarski2011}. Its soft state is peculiar and sometimes called intermediate \citep{belloni1996}. These differences may rely on the high mass nature of Cyg X-1 system, where wind fed accretion takes place, as compared to the vast majority of other BHBs which host low mass companions.

 Cyg X-1 fast variability features were well studied by e.g. \cite{pottschmidt2003}, making extensive use of the huge {\it RXTE} data archive. They confirmed that the PDS could be well fitted by sets of Lorentzians, and they showed that the relative contributions of those sub-components were evolving with the states. The lower frequency Lorentzian is indeed suppressed relative to the second and third Lorentzian during the state transitions. Its energy behavior, or rms-spectrum, is rather constant in the hard state at $\sim 25-40\rm \%\ rms$.

There is however an important question which remains to address: are these high frequency variability features still present in the hard X-rays? At higher energies, a strong LFQPO (Low Frequency QPO) was indeed detected by SIGMA on board GRANAT \citep{vikhlinin1994} in March 1990 and March 1991. Its typical peak frequency (between 0.04 and 0.07 Hz) was found to be similar to the broad band noise break frequency that was also detected in the power spectrum. However its behavior was very variable as the QPO, together with the broad band noise had disappeared one year later.

 The energy dependency of the variability was also extensively studied in other sources: \cite{rodriguez2004d, rodriguez2004c} found that a cut-off near 25 keV was probably present in GRS 1915+105 and XTE J1550-564 for the LFQPO rms-spectrum. Concerning the long time scales variability (from days to weeks), \cite{zdziarski2002} used the {\it RXTE}/ASM and CGRO/BATSE archives to study the variability in Cyg X-1 from 1.2 to 300 keV. They showed that this very low frequency variability was also quite high (i.e. 25\% rms) and state dependent. Examining Cyg X-1 fast variability as a function of state, \cite{gierlinski2010} (hereafter GZD10) showed that the rms spectrum was almost flat for the hard state whereas it is increasing with energy in the soft state. In intermediate states, it is decreasing with energy. Similar trend is observed in other BHBs (see e.g. \citealp{gierlinski2005}). A tight and strong correlation was also found between the Comptonization amplification factor ($\ell_{\rm h}/\ell_{\rm s}$) and the frequency of the first Lorentzian $f_1$ as $\ell_{\rm h}/\ell_{\rm s}\propto f_1^{-3/2}$.

The models to explain those variability features are still under debate, principally due to the various patterns observed and their evolution with the state. Its origin, i.e., is it coming from the disk, the corona, from a transition region or from the jet is even not well established. The most probable hypothesis is that it relies in an interaction between those components. Some successful attempts have however been built in order to reproduce some of the features observed: Concerning e.g. the LFQPO origin, the Accretion-Ejection Instability \citep{tagger1999} was used  to explain the LFQPO frequencies and the disk geometry evolution \citep{varniere2002}. \cite{cabanac2010} built a model of an oscillating corona which addresses the QPO and the broad band noise typical frequencies correlation, the modeling of the PDS by radiative transfer simulation as well as the qualitative evolution of the frequencies with states : the softer the SED, the higher the frequencies. The observed increase of the absolute variability with the source flux, so called ``RMS-flux correlation'' \citep{uttley2005}, was finally well modeled by \cite{arevalo2006}, in the context of fluctuating-accretion model.

The purpose of this article is to study the fast timing behavior of Cyg X-1 at higher energies by using the high sensitivity of SPI (Spectrometer Onboard Integral, \citealp{vedrenne2003,roques2003}), as its average effective area is $\sim 500\ \rm cm^2$ in the 20 keV - 8 MeV band. Its timing ability is indeed unprecedented in those high energy bands, and it allowed, e.g., absolute timing of the Crab and determination of the hard X-rays to radio delay at  $\sim 15 \rm \mu s$ resolution \citep{molkov2010}.
\section{Data analysis Principles}
\subsection{Event extraction}\label{subsec:event-extract}
A data analysis pipeline was build for timing analysis purpose of SPI data. This set of Perl scripts is designed to first extract the event files, and second the energy spectra associated with given sources in the FOV. 
The source spectral extraction is needed for evaluation of the source and background contribution to the variability, see hereafter. 
 For that purpose, we used our local {\verb deconv }\ algorithm in our pipeline (see e.g. a description of the principles in section 2.3 of \citealp{jourdain2009}).
The event extraction process is in turn based on the algorithm used in \cite{molkov2010}. 
Some caveats described just below were moreover taken into account.

First, in order to limit contribution from other sources, we only retained the pointed observations where Cyg X-1 was in the Fully Coded Field Of View (hereafter FCFOV) of SPI, i.e., when the instrument axis is closer than 7$^{\rm o}$ from the source. Such a choice led us to retain about 1000 SCWs.
 
Some SCWs are however affected by telemetry gaps in the satellite to earth data transmission. As these lost packets are present in less than 10\% of the observations, and since it affects the Fourier analysis, discarding them was preferred. 
 Furthermore, only single events were retained for our study \citep{roques2003}. As stated in \cite{molkov2010}, this choice was motivated by the fact that the timing resolution is worse with multiple events, together with the difficulty of spatial information retrieving. Barcentric correction of the event files were performed using the standard ISDC {\verb barycent } tool.

Finally, two energy bands relevant for our study were selected: the lowest energies were probed between 27 and 49 keV, whereas the high one is a gather of the 69-90 and 96-130 keV bands (abbreviated as 69-130 keV hereafter). Such choice is motivated for increasing the source to background ratio in the event files by avoiding instrumental lines (see e.g. \citealp{weidenspointner2003} for the identification and the modeling of those lines).
   
Note that the previous caveats are automatically taken into account by our SPI timing pipeline, and unless explicitly mentioned, every subsequent errors tabulated or in the text are at 90\%. By default, the plotted ones are at $1\sigma$. The latter confidence levels for a given $value$ in this article will be expressed as $value_{min\ value}^{max\ value}$. 
\subsection{Evaluation of the noise level in the PDS}\label{noise-level}
Before calculating any power spectra for a given source, it is necessary to check what is the shape and the level of the PDSs in absence of the source. For background dominated instruments, the count rate is high, even when observing an empty field region. Hence it is possible to evaluate the noise level with good statistics if empty field observations are used. We therefore used 45 SCWs from rev. \# 681, corresponding to a total of $\sim 156\ \rm ks$, when {\it INTEGRAL} was pointing towards mid galactic latitude (RA=$19^h42'44.23''$, Dec=$-15^h04'03.4''$) , and no significant solar activity was observed. In order to know how close the noise behaves like Poisson statistics, an average empty field PDS (hereafter EFPDS) of rev \# 681 was calculated for both 27-49 keV and 69-130 keV bands. The obtained EFPDS was normalized to Leahy \citep{leahy1983a}, such that the comparison with theoretical behavior for Poisson statistics is easier. In that latter case, the obtained PDS must be flat when averaged-up on infinite number of time frames and a value of 2 is expected for purely Poisson noise in Leahy normalization \citep{leahy1983a}.

\begin{figure}[t]
\centering
\includegraphics[angle=-90,scale=0.4]{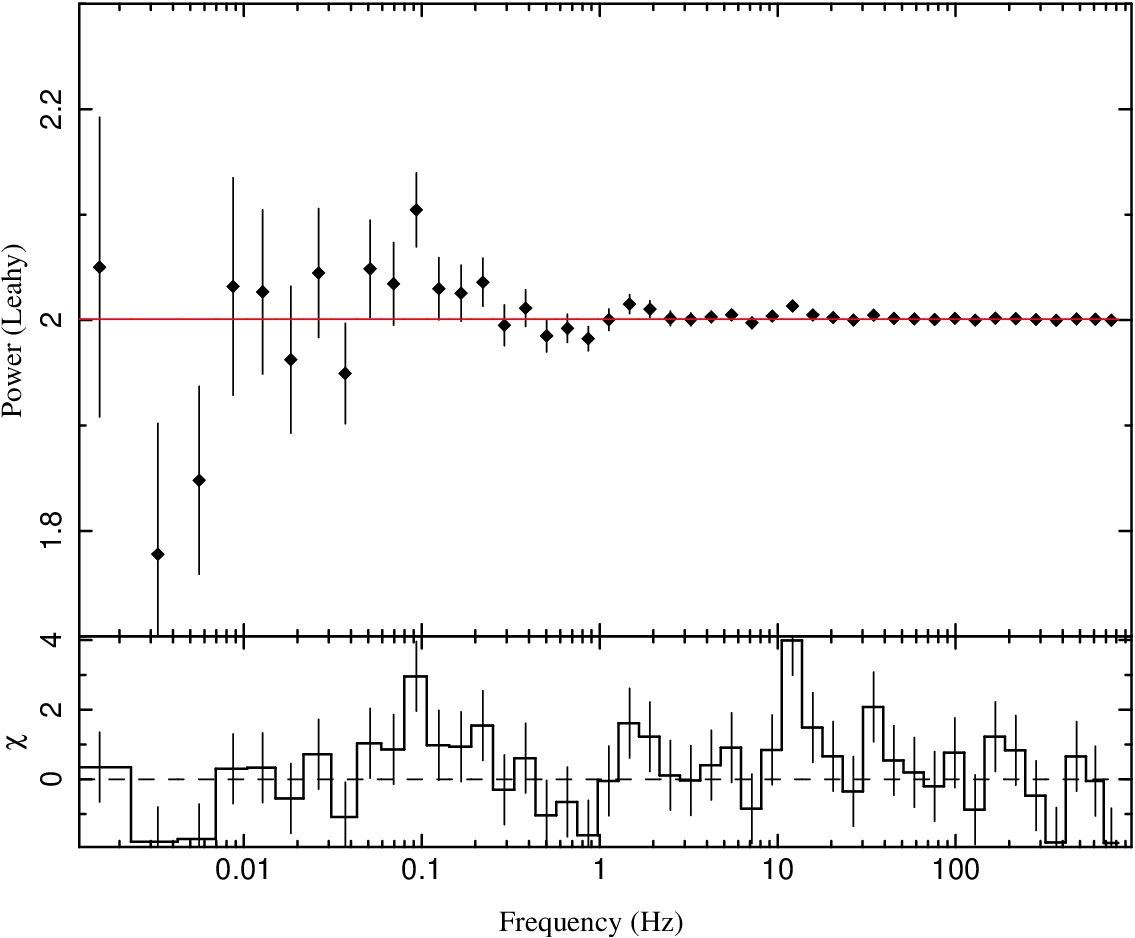}
\includegraphics[angle=-90,scale=0.4]{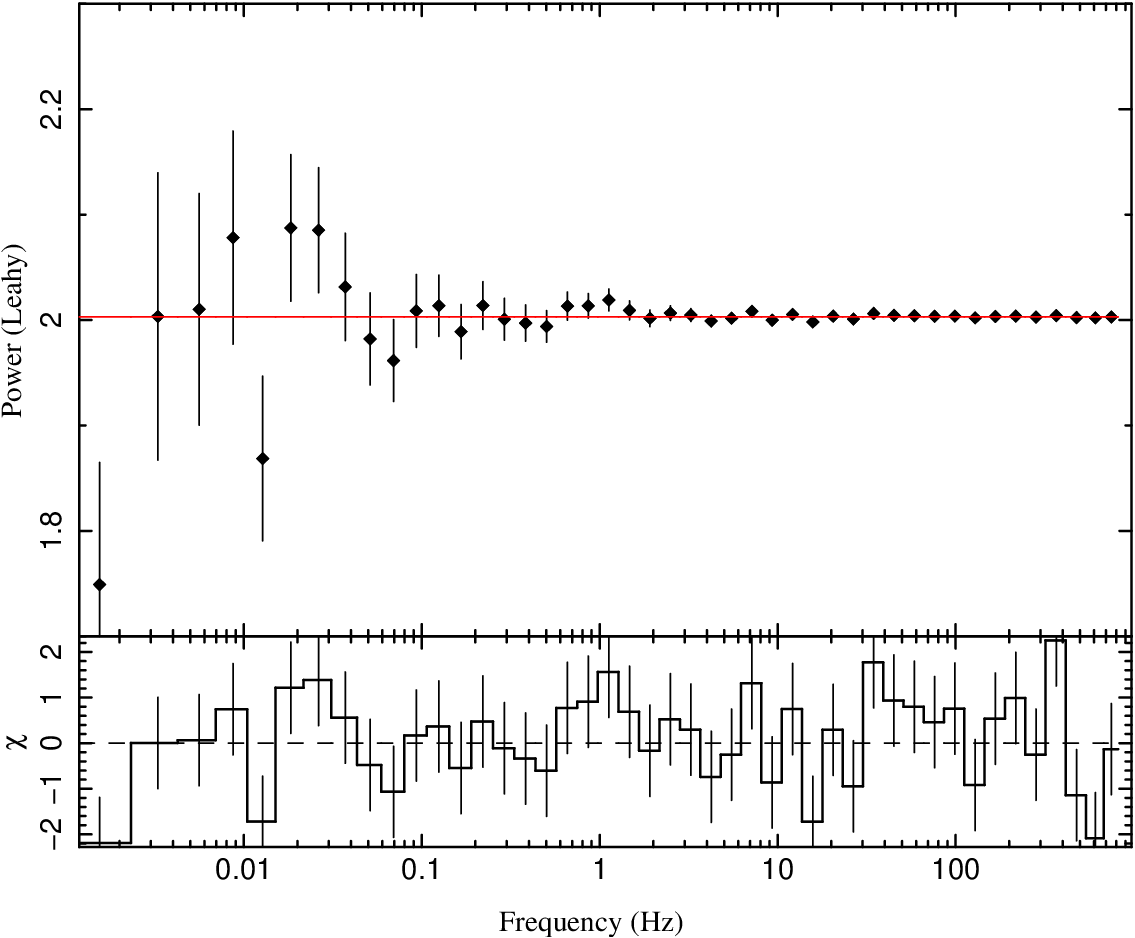}
\caption{Power spectra of the empty field observed during revolution \#681 in the 27-49 keV band (upper panel) and the 69-130 keV band (lower panel). Note that except for few points, the PDSs are well fit by a constant (solid horizontal line, see text for the parameter of the fits) .\label{fig:FFpsd}}
\end{figure}
For the  27-49 keV band (see upper panel of Fig. \ref{fig:FFpsd}) an average value consistent with 2 was obtained. More precisely, the result of the fit indeed gives $P_{FF,\ 27-49\ keV}=2.00077^{2.00109}_{2.00045}$.  The statistic for this fit is $\chi^2/\nu=1.56$ for 44 dof, whereas it is $\chi^2/\nu=1.88$ when the noise level is set to the theoretical value of 2. The moderately good $\chi^2$ obtained after fitting is due to the presence of single peaks in frequency bins, but not to an overall trend (e.g. deviation in slope), as either low or high frequency bins contribute equally to the $\chi^2$. Hence, it is possible that the peaks observed for the 27-49 keV are only due to statistic fluctuations (since our empty field data sample is not infinite), but we cannot exclude either instrumental, or more probably environmental effects.
 
Concerning the 69-130 keV band, it is still consistent with a constant close to 2 (see lower panel of Fig. \ref{fig:FFpsd}), but again the fitted value is slightly higher as $P_{FF,\ 69-130\ keV}=2.0030^{2.0033}_{2.0026}$. The statistics is however quite good as $\chi^2/\nu=1.03$ for 44 dof, with no significant deviations to the constant obtained. When the noise level is set to 2, the statistics give $\chi^2/\nu=6.16$. 

We tentitatively looked for the reasons of such deviations to the pure Poisson noise process, the main one being the possible effect of deadtimes on the obtained PDSs. Modeling of such effects were performed by e.g. \cite{zhang1995} and used with success on the fine noise level  evaluation of {\it RXTE}/PCA PDSs (see e.g. \citealp{morgan1997}). In practice, we used the formula given in \cite{nowak1999} and applied it to SPI detector typical values. We obtained a theoretical Poisson noise level spanning between $1.98924$ and $1.99017$ in Leahy normalization. This theoretical level is therefore very different from the value obtained when computing the actual EFPDS, and it is possible that some other instrumental effects than deadtimes are affecting the EFPDS. Another possibility of this excess could be the presence of weak sources in the ``empty'' FOV chosen, which may increase the theoretical noise level.
Since such deviation from the theoretical noise level may also affect the subsequent source PDS calculation and its rms evaluation, we built a practical method to overcome this effect: even with instrumental effects indeed, the source PDS always converge to the noise level at sufficient high frequencies. Hence, we determined the empirical noise level in each PDS by setting an arbitrary frequency threshold $f_{min}$, and then fit the PDS over $f \geq f_{min}$ with a constant $C_{noise}$, which gives finally our practical noise level. Naturally, the noise level obtained may contain some source intrinsic variability as $f_{min}\neq\infty$. Hence, some high frequency components can be missed or artificially added with that method. We note that a  similar method (i.e., fit of the  noise level at high frequencies) was used with success on {\it RXTE} data for the search of twin kHz QPO in neutron stars \citep{boutelier2010},  and we examined the possible effects on an example below (in section \ref{subsec:301-684}).
\subsection{Background dominated PDS}\label{subsec:back-dom-psd}
In the case of background dominated data, corrections have to be applied in order to evaluate the intrinsic power spectrum of the source. We indeed used the raw count rate detector lightcurve as input for calculating the raw power spectrum $P_{tot,\ Leahy}$. It is then easy to demonstrate that the intrinsic power spectrum $P_{src,\, rms}$ expressed in ``rms'' or ``Myiamoto'' normalization (\citealp{belloni1990},  or \citealp{miyamoto1991}) follows:
\begin{equation}
P_{src,\, rms}= \left[P_{tot,\, Leahy}-C_{noise}\right]\frac{\langle B+S\rangle}{\langle S\rangle^2},
\end{equation}
where $\langle S\rangle$ is the average source count rate, and $\langle B+S\rangle$ the average detector count rate (see also \citealp{vaughan1997}). For coded mask instruments, the value of $\langle S\rangle$ is obtained by using the usual deconvolution algorithms for spectral extraction, and the average detector count rate $\langle B+S\rangle$ is directly given by the event files.
\section{Results}
 For the power spectra computation, we arbitrarily set each segment duration to $\Delta T=647\ \rm s$, fixed the binning time such that the highest frequency available is $f_{Nyq}=810\ \rm Hz$ and rebinned logarithmmically the power spectra obtained. For that purpose, we used the standard {\verb powspec }\ XRONOS tool to compute the mean power spectrum $\langle P_{tot,\, Leahy}\rangle$.
We then used {\verb ISIS }\footnote{\url{See http://space.mit.edu/CXC/isis/index.html}} version 1.6.1-26 \citep{houck2000}, as well as routines publicly available from the S-lang/ISIS Timing Analysis ({\verb SITAR })\footnote{\url{http://space.mit.edu/CXC/analysis/SITAR/}} for PDSs post-processing (fitting, time lags and rms evaluation).
\begin{figure}[t]
\centering
\includegraphics[angle=-90,scale=0.4]{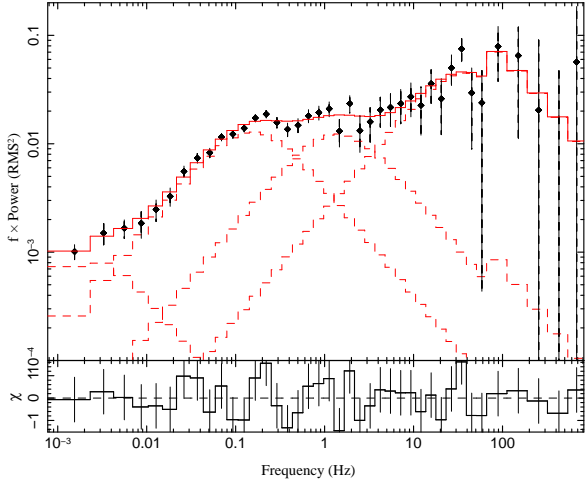}
\caption{Power spectrum of Cyg X-1 calculated from rev \#301 to 684 (observations where EXO 2030+375 was flaring were removed) in the 27-49 keV band. The intrinsic average fractional variability is $rms\sim 29\pm 3\%$. 4 Lorentzians were used to fit the spectrum. See first column of table \ref{tbl:301-684-rms} for the parameters of the fit.}\label{fig:psd-301-684}
\end{figure}
The noise level was finally evaluated as prescribed in the previous section, i.e., we fitted the high frequency tail of each PDS by a constant. The minimum frequency chosen for the fit of $C_{noise}$ was set to $f_{min,\ noise}=100\ \rm Hz$. We also rebinned again the obtained PDSs at high frequencies for better statistics which explains the non equal bin spacings in the PDS plotted.
\begin{figure}[t]
\centering
\includegraphics[angle=-90,scale=0.4]{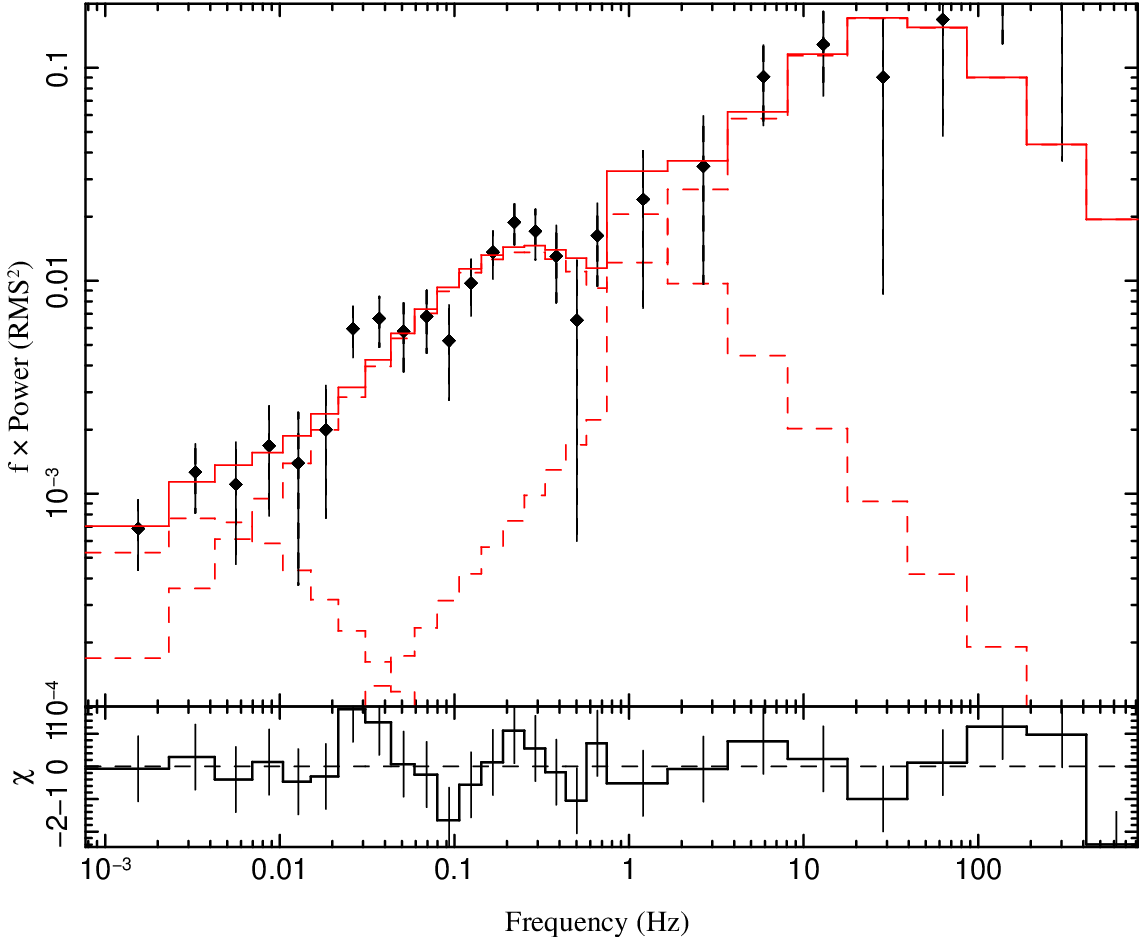}
\includegraphics[angle=-90,scale=0.4]{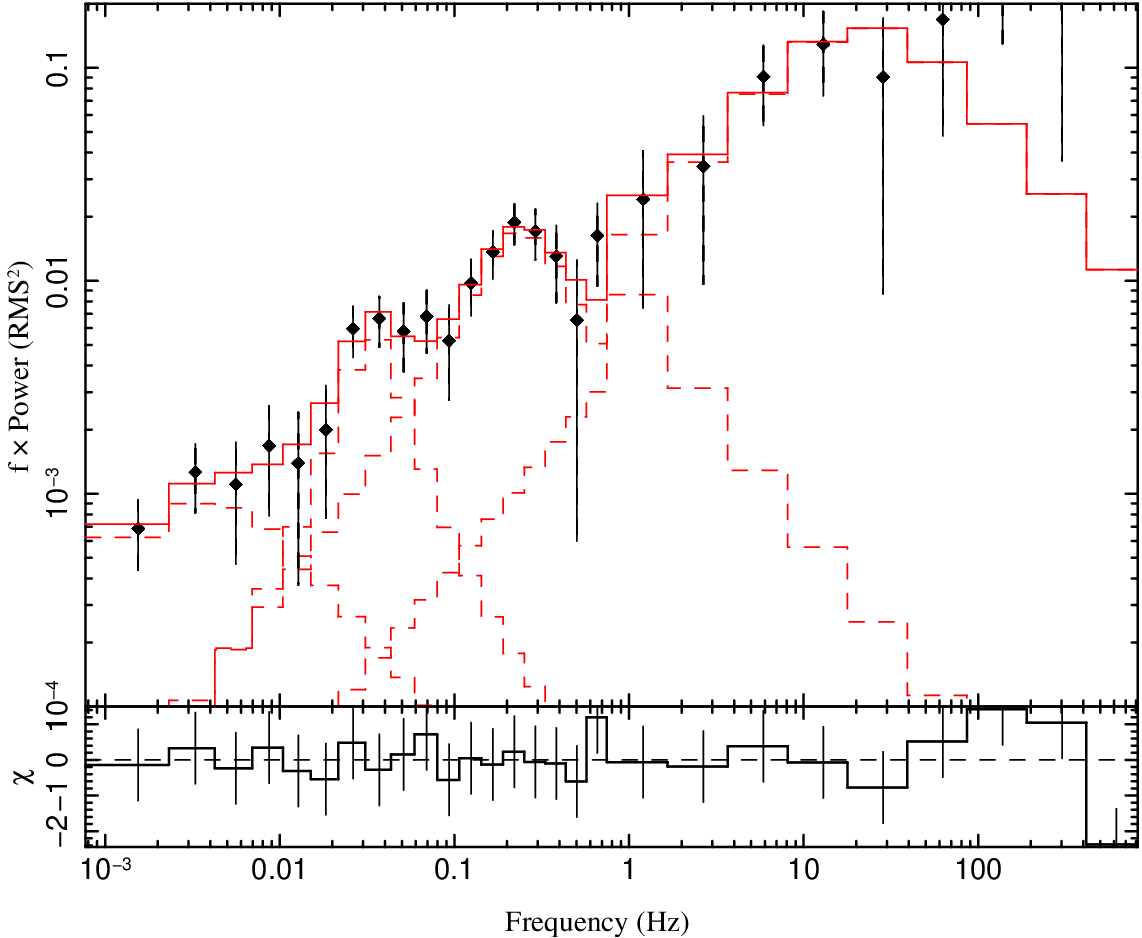}
\caption{Power spectrum of Cyg X-1 calculated from rev \#301 to 684 (observations where EXO 2030+375 was flaring were removed) in the 69-130 keV band. (Up): 3 zero centered Lorentzians were used to fit the PDS. (Down): 2 zero centered and 2 ``QPOs''. See second column of table \ref{tbl:301-684-rms} for the parameters of the fit.}\label{fig:psd-301-684-69-130}
\end{figure}
\subsection{State averaged power spectra of Cygnus X-1}\label{subsec:301-684}
We first focused in our study on state-averaged power spectra computation. The following PDS hence mixed different physical state of the source, but this choice was motivated primarily for increasing the statistic at high frequencies in the highest energy band probed. We also found that averaging up the observation from revolutions \#301 to 684 were the best trade off for increasing the statistic in the 69-130 keV bands. Earlier observations (especially revolutions \#79 and 80) are indeed quite soft in the SPI energy range, and therefore decreases the signal to background ratio of the source in that band.

 Noise subtracted power spectra of Cyg X-1 in ``rms'' normalization are plotted in Fig. \ref{fig:psd-301-684} for the 27-49 keV band and in Fig. \ref{fig:psd-301-684-69-130} for the 69-130 keV band. They were calculated from rev \#301 to 684, but during this period the accreting pulsar EXO 2030+375 located $\rm 7.2^{o}$ from Cyg X-1 was flaring. Hence, as the total SPI FOV is $\rm 35^{o}$ ($\rm 16^{o}$ for the fully coded FOV), and to prevent any contribution in the calculated PDS from that source, we removed from our data sample the observations corresponding to the dates of the flare, i.e from rev \#452 to 486. Disentangling any contribution from this source in the PDS is the purpose of the appendix.

Our first result is that the overall shape of the noise subtracted PDSs in the 27-49 keV band shown on Fig. \ref{fig:psd-301-684} looks very similar to the ones obtained with softer X-ray monitoring of the source by e.g. {\it RXTE} (see e.g. \citealp{lin2000, nowak1999}). Fitting the obtained PDS by a constant gives in turn a very bad reduced $\chi^2$ since $\chi^2/\nu=37.8$ for 44 dof.
We therefore chose to fit the PDSs as it is usually done in the literature, i.e., with sets of Lorentzians $L_i$ \citep{nowak2000, belloni2002, pottschmidt2003}, i.e.;
\begin{equation}
L_i(\nu) = \frac{1}{\pi \Delta f_i}\frac{2R_i^2}{1 + 4 \left(\frac{\nu - f_i}{\Delta f_i}\right)^2},
\end{equation}
with $f_i$ the resonance frequency of the Lorentzian, $\Delta f_i$ the full frequency width at half maximum value, and $R_i$ a normalization constant. We then define the quality factor $Q_i$ of this Lorentzian as:
\begin{equation}
Q_i=\frac{f_i}{\Delta f_i}.
\end{equation}
 the normalization $R_i$ is linked to the rms of this Lorentzian by integrating $L_i$ over the whole frequency domain, which gives (see also \citealp{pottschmidt2003}):
\begin{equation}
rms_i=R_i\sqrt{\frac{1}{2} + \frac{atan(2Q_i)}{\pi}}
\end{equation}
In the following text, fitting the PDS with ``QPO'' refers to Lorentzian with $f_i \neq 0$. However, for sake of simplicity, we generally used by default zero frequency Lorentzians, i.e., $f_i = 0$. This choice was motivated for decreasing the number of free parameters in the fit while giving reasonable $\chi^2$ values. In that case, $Q_i=0$ and the only relevant parameters are $\Delta f_i$ and the normalization $R_i=rms_i\sqrt{2}$.

For example,  fitting the 27-49 keV PDS with only three Lorentzians gives a reduced  $\chi^2/\nu=1.53$ for $34$ degrees of freedom (dof), leaving a high frequency tail unfitted. The addition of a fourth zero centered Lorentzian gives in turn  $\chi^2/\nu=0.72$ for 32 dof. It corresponds to a F-test Null Hypothesis Probability (hereafter NHP) of $2.4\ 10^{-6}$ for this addition (see however \citealp{protassov2002} for the limited uses of the F-test), which gives good confidence in the presence of such fourth component. However, since the presence or the absence of this fourth component may be affected by an under estimation of the noise level, we also reanalyzed the data in a worse case: we set the minimum frequency for the noise level fit to $f_{min,\ noise}=0.005\ \rm Hz$ instead of $100\ \rm Hz$. With that choice, the noise level obtained will always increase, as it may include some source contribution as well. Therefore, after subtracting this new noise level, we re-fitted the PDS with three or four Lorentzians. The obtained $\chi^2$ values are naturally worse, as compared to the previous case where $f_{min,\ noise}=100\ \rm Hz$, as they are dominated by systematic negative residuals at high frequencies: we obtain $\chi^2/\nu=1.83$ with 34 dof for 3 Lorentzians and $\chi^2/\nu=1.29$ with 32 dof for 4 Lorentzians. Hence, even with this extreme evaluation of the noise level, the F-test NHP for the addition of the fourth Lorentzians is equal to $1.5\ 10^{-3}$. This gives in turn good confidence for the presence of such high frequency tail in our data set.

The fit parameter values, together with their corresponding errors are displayed in Table \ref{tbl:301-684-rms}. The errors on each fitted parameter were calculated leaving all the others free to vary, unless mentioned explicitly. The four fitted Lorentzians frequencies values span from $2. 10^{-3}$ to $35\ \rm Hz$ and are logarithmically spaced by a factor $\sim 5$ to $20$. Besides the very low frequency component of frequency $\Delta f_0$ which account for less than 3\% of the variability, the low frequency Lorentzians ($L_1$ and $L_2$) contribute to the total rms by the same order of magnitude amplitude, i.e., 12-13\%, and it increases to 23\% for the high frequency ($L_3$) Lorentzian.  The overall Cyg X-1 variability in the 27-49 keV band is therefore estimated to $29.4\pm 2.8\%$ rms from revs \#301 to 684.    

  \begin{deluxetable}{c|c||c c}
\tabletypesize{\scriptsize}
\tablecaption{Revs \#301 to 684: results of the PDS fit by 3 or 4 zero centered Lorentzians. \label{tbl:301-684-rms}}
\tablewidth{0pt}
\tablehead{
 \colhead{Energy band}                & \colhead{27-49 keV} & \multicolumn{2}{c}{69-130 keV}
}
\startdata
                 &                   &  \tablenotemark{(a)}                              &  \tablenotemark{(b)}\\
          $\displaystyle \Delta f_0$ (Hz)    &  $\displaystyle  0.0024^{  0.0043}_{ 0.00087  }$   &  $\displaystyle 0.0038_{     0.0002     }^{    0.038 }$         & $\displaystyle  0.0024\ (f) 			$           \\
        $\displaystyle rms_0$         &     $\displaystyle 0.032^{ 0.037}_{  0.026}$     &      $\displaystyle 0.031_{ 0.019    }^{   0.053 } $          & $\displaystyle  0.027^{   0.035   }_{    0.016 } 	$   \\
         $\displaystyle  \Delta f_1 $(Hz)    &    $\displaystyle  0.15^{  0.18}_{   0.13}  $    &     $\displaystyle 0.25_{  0.17 }^{       0.36 }$            & $\displaystyle   0.15\ (f) 			$           \\
       $\displaystyle  rms_1$         &    $\displaystyle  0.13^{  0.14}_{   0.12}$      &       $ \displaystyle  0.13_{ 0.12    }^{    0.15 }$           & $ \displaystyle    0.11  ^{    0.12  }_{    0.10 } 	$   \\
        $\displaystyle   \Delta f_2$ (Hz)    &    $\displaystyle 1.39^{2.34}_{0.85}$             &  \nodata	    & $ \displaystyle   1.39\ (f) 			$           \\
       $\displaystyle  rms_2$         &    $\displaystyle  0.12^{  0.14}_{0.11}$         &   \nodata      & $ \displaystyle     0.09  ^{     0.13	}_{     0.0 } 	$   \\
        $\displaystyle   \Delta f_3$ (Hz)    &    $\displaystyle 38.1^{57.8}_{22.3}$             &  $ \displaystyle  32^{ 155    }_{  9   }$ 	    & $ \displaystyle      38.1\ (f) 			$           \\
       $\displaystyle  rms_3$         &    $\displaystyle  0.23^{  0.27}_{0.20}$         &    $\displaystyle  0.26^{  0.40}_{0.17}$           & $ \displaystyle     0.26  ^{     0.32	}_{     0.19 } 	$   \\
$ \displaystyle  rms_{tot}/mean$                           & $\displaystyle 0.29^{0.32}_{0.27}$                & $ \displaystyle 0.30_{0.23}^{0.36}$             &  $ \displaystyle  0.30^{0.35}_{0.21}$                 \\ 
$\displaystyle  \chi^2/\nu$           &      0.87                                         & 1.04                                            & 1.08                                  
\enddata
\tablecomments{$(f)$ stands for a parameter value fixed in the fit, $(p)$ for a parameter that pegged at a limit during the error calculation.}
\tablenotetext{(a)}{Values obtained when all parameters of the fit are let free to vary.}
\tablenotetext{(b)}{Values obtained when tying the  Lorentzian frequencies to the value obtained in the 27-49 keV band.}
\end{deluxetable}
For the 69-130 keV band, the PDS qualitative shape was showing even more deviation to the powerlaw shape with clear defined ``humps'' (see Fig. \ref{fig:psd-301-684-69-130}). Again, the PDS fitted with a simple constant gives $\chi^2/\nu=5.26$ for 44 dof. When fitted with a model of 3 zero-centered Lorentzians, a quite good reduced $\chi^2/\nu=1.04$ for 31 dof is obtained. The addition of a fourth Lorentzian is not required as it gives roughly the  same $\chi^2$, and the corresponding F-test NHP is in turn $38\%$. Note that the Lorentzian which is not required as compared to the 27-49 keV is the ``mid'' frequency ones (with frequency $\Delta f_2$). Since the errors on the parameters are bigger than the ones obtained for the 27-49 keV,  we also tried to fit the 69-130 keV PDS by tying the typical frequencies of the Lorentzians to the values obtained at low energies. This in turn gave a similar  $\chi^2$ value ($\chi^2/\nu=1.08$ for 34 dof), and the total rms values are equivalent, typically around $30\%$ rms. It is also interesting to note that when the frequencies are tied in the 69-130 keV fit, the rms of each sub-component is consistent with the one found for the 27-49 keV band.


As the the 69-130 keV PDS exhibits humps and especially at low frequencies, fitting  them with non zero centered Lorentzians (i.e., QPOs) was also tried (see lower panel of Fig. \ref{fig:psd-301-684-69-130}). The reduced $\chi^2$ value can be reduced down to $0.80$ for $17$ dof by this way, if a model composed by 2 zero centered Lorentzians and 2 ``QPOs'' is used. This alternative model does not affect the overall variability estimation as in that case $rms_{tot}/mean=0.28^{0.33}_{0.24}$. The values of the quality factor $Q_i$ are  $Q_1=1.24^{1000}_{0.17}$ and $Q_2={0.62^{1.4}_{0.01}}$ for both ``QPOs'' fitted. Hence, since $Q<2$, those humps found in the 69-130 keV PDS may be qualified as ``peaked noise''  (see however e.g. \citealp{belloni2002} and reference therein for a  criticism about such a strict QPO/peaked-noise boundary). The frequencies of both peaked noise are in turn $f_1=0.033^{0.040}_{0.021}\ \rm Hz$ and $f_2=0.22^{0.25}_{0.14}\ \rm Hz$.
The results of the 69-130 keV PDS zero centered Lorentzians fits are given in Table \ref{tbl:301-684-rms}. In summary, the PDSs shows strong variability from $10^{-3}$ to at least a few tens of Hz, reaching a level of roughly 30\% rms, even at those high energies. The low frequency component ($L_1$ or $L_2$ when present) contributes for $10-15\%$ to the rms. In contrast, the high frequency tail contributes for at least $26\%$. 

\subsection{Evolution of the rms}
We estimated our overall rms by integrating each modelled PDS. In this section, we studied the evolution of the variability and its dependence with the state or the energy. 
\subsubsection{RMS as a function of time}
 The evolution of the variability as a function of time was first examined. In order to obtain sufficient statistics for evaluating the rms, but also its evolution with time, a one revolution typical time scale was found to be the best trade-off for grouping the data together. For the same reasons, we also excluded the revolutions where Cyg X-1 was in the FCFOV during less than 20 SCWs. Hence, 16 different revolutions were selected, spanning from revs \#79 (June 2003) to 684 (May 2008), and corresponding to a total of $2.2 \rm\ Ms$ of data. Noise subtracted PDSs were then computed for each revolution in the same way as done in section \ref{subsec:301-684}. Only 1 or 2 zero centered Lorentzians were necessary to fit each of them well, given the lower statistics compared to the previous grouped analysis.
\begin{figure*}
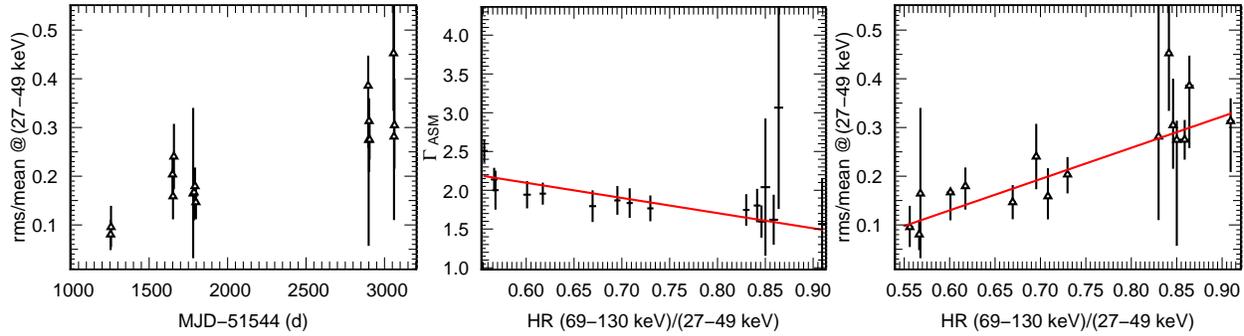

\centering
  \includegraphics[angle=0,width=0.3\textwidth]{rms_vs_time.epsi}
  \includegraphics[angle=0,width=0.3\textwidth]{HR_spi_vs_gamma_asm.epsi}
\includegraphics[angle=0,width=0.3\textwidth]{rms_vs_HR_fit.epsi}
  \caption{Evolution of Cyg X-1 timing and spectral parameters from our sample. {\it Left panel:} Evolution of $rms_{tot}/mean$ in 27-49 keV band with time.  {\it Center panel:} SPI hardness ratio (HR) as a function of the derived ASM photon index $\Gamma_{ASM}$. {\it Right panel:} rms vs SPI HR. Straight lines show the fits of the points in the middle and right plots by a linear function. 90\% errors plotted.}\label{fig:rms-flux-hr-vs-time}
\end{figure*}
In order to compare the timing parameters evolution with the spectral ones, we also built deconvolved spectra for each selected revolution. Using those spectra, the source count rates in the 27-49 keV and the 69-130 keV bands were finally computed. In order to get a quick look on the spectral shape of the source, we also defined a SPI hardness ratio as:
\begin{equation}
\displaystyle HR_{SPI}= \frac{Count\ Rate_{\ 69-130\ keV}}{Count\ Rate_{\ 27-49\ keV}}.
\end{equation}
The rms in the 27-49 keV band as a function of time is shown on Fig. \ref{fig:rms-flux-hr-vs-time}. The overall trend is that the total fractional rms evolves with time from about $10\%$ to $35\%$. This variable behavior led us to examine any possible correlation between the calculated rms and the spectral parameters.  
\subsubsection{RMS - hardness ratio correlations}
The fractional rms as a function of  SPI hardness ratio is plotted in Fig. \ref{fig:rms-flux-hr-vs-time}. The overall clear trend is that the fractional variability tends to correlate with $HR_{SPI}$: rank coefficients were computed  (see Table \ref{tbl:correlation}) and gave a NHP equal $2. 10^{-5}$ ($2. 10^{-4}$ for the Kendall's test), hence the rms is well correlated with $HR_{SPI}$. 
$\chi^2$ fitting of the previous relations were also tried, in order to take errors into account. Simple linear functions were used for those fits (straight lines in Fig \ref{fig:rms-flux-hr-vs-time}) and the correlation between the rms and $HR_{SPI}$ is still confirmed by this way. The line slope is indeed $S_{rms-HR_{SPI}}=0.64\pm 0.11\ \rm rms$. 

\begin{deluxetable}{c|c|c||c|c}
\tabletypesize{\scriptsize}
\tablecaption{Results of the correlation tests between the fractional $rms$ in the 27-49 keV band and count rates or HRs. Prob. refers to the null hypothesis probability under the considered test (Spearman's $\rho$ or Kendall's $\tau$ rank coefficients). \label{tbl:correlation}}
\tablewidth{0pt}
\tablehead{
\colhead{$\% rms_{tot}$ vs/} &\colhead{$\rho$} & \colhead{Prob.} & \colhead{$\tau$} & \colhead{Prob.} 
}
\startdata
SPI count rate    & -0.31 & 0.23           & -0.22 &  0.24             \\
SPI HR      & 0.85  & $\displaystyle 2.0\ 10^{-5}$ &  0.68 &  $\displaystyle 2.2\ 10^{-4}$   \\
ASM count rate    & -0.71 & $\displaystyle 1.9\ 10^{-3}$ & -0.48 &  $\displaystyle 9.0\ 10^{-3}$             \\
ASM soft HR (B/A)& 0.87  & $\displaystyle 1.1\ 10^{-5}$ &  0.73 &  $\displaystyle 7.4\ 10^{-5}$   \\
ASM hard HR (C/B)& 0.49  & 0.05           & 0.40  & 0.03              
\enddata
\end{deluxetable}
\subsubsection{ASM photon index vs SPI hardness ratio}
The important question of how $HR_{SPI}$ relates with the state definition usually based on soft X-ray criteria was then examined. For that purpose, we retrieved from the {\it RXTE}-ASM archive\footnote{\url{http://xte.mit.edu/asmlc/srcs/cygx1.html}} the simultaneous count rates of the source in the low (A, 1.5-3 keV), medium (B, 3-5 keV) and high (C, 5-12 keV) standard energy bands. Those count rates were then converted into fluxes and 3-12 keV photon indices ($\Gamma_{ASM}$) of the source, following the method described in \cite{zdziarski2002}. Finally, $HR_{SPI}$ was plotted against the obtained $\Gamma_{ASM}$ (see center Fig. \ref{fig:rms-flux-hr-vs-time}). It appears that $HR_{SPI}$ is anti-correlated with $\Gamma_{ASM}$ (Spearman's rank coefficient is here $\rho=-0.49$), as when fitted with a straight line, the $HR_{SPI}$ vs $\Gamma_{ASM}$ gives a negative slope of $S_{HR_{SPI}\ vs\ \Gamma_{ASM}}=-1.95^{-1.19}_{-2.72}$. The y axis intercept value for this straight line is in turn $\Gamma_{HR_{SPI}=0} = 3.27^{3.78}_{2.75}$.
\subsection{Variability in Hard State vs Soft State}
In the following section we examined the fractional variability of Cyg X-1 as a function of state. For defining those states, we used two different methods. The first is based on the photon index $\Gamma_{ASM}$ obtained in soft X-rays, and the second one is based on the hardness ratio obtained in the hard X-ray band ($HR_{SPI}$). Both are anti-correlated as demonstrated in the above paragraph. However, due to the scatter of $\Gamma_{ASM}$ vs $HR_{SPI}$ plot, grouping of the observation based on one or the other criterion lead to different samples, and hence, as we will demonstrate, different results.
\subsubsection{States defined by the ASM} 
\begin{figure*}
\centering
\includegraphics[angle=-90,scale=0.4]{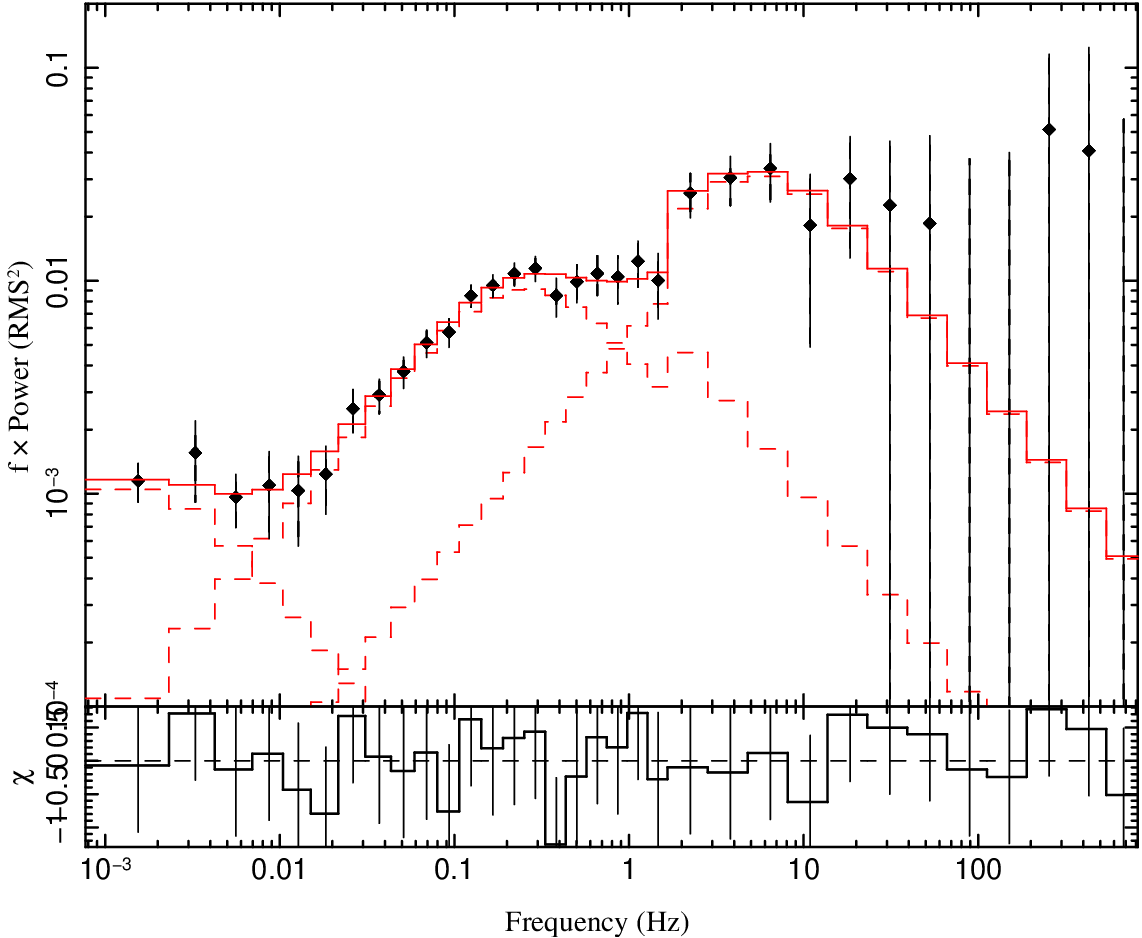}
\includegraphics[angle=-90,scale=0.4]{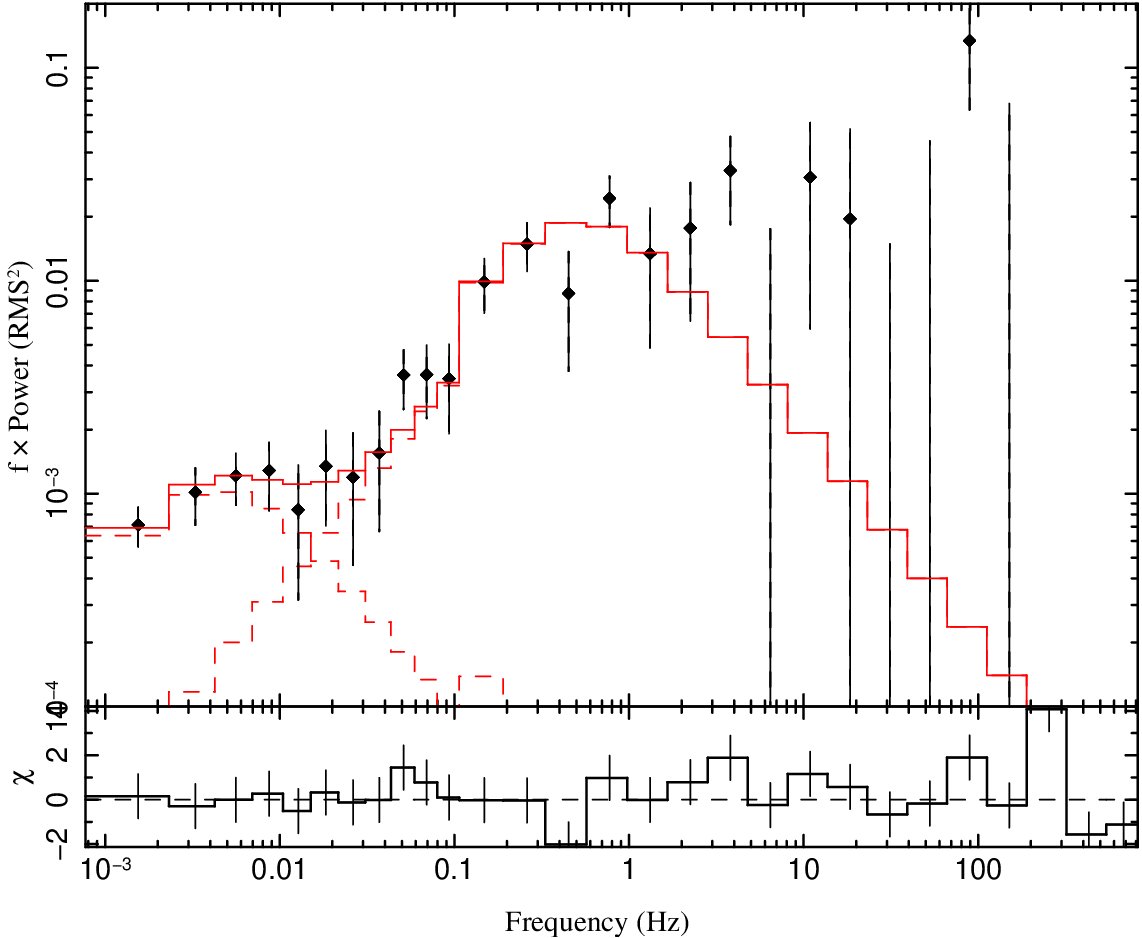}
\caption{Cyg X-1 PDSs calculated when the source was hard (left panel) and soft (right panel) according to the ASM criterion in the 27-49 keV band.}\label{fig:psd-hard-soft-asm}
\end{figure*}

The first state definition is based on the derived value of $\Gamma_{ASM}$ by:
\begin{itemize}
\item $\Gamma_{ASM}<2$ corresponds to a hard state ($HS_{ASM}$), whereas,
\item $\Gamma_{ASM}\geq 2$ corresponds to a soft state ($SS_{ASM}$).
\end{itemize}
The same process as described in section \ref{subsec:301-684} was applied to the grouped $HS_{ASM}$ and $SS_{ASM}$ event files. The 27-49 keV PDSs are plotted in Fig. \ref{fig:psd-hard-soft-asm}. 

For the $HS_{ASM}$, the 27-49 keV PDS is remarkably well fit by three Lorentzians ($\chi^2/\nu=0.29$, whereas $\chi^2/\nu=1.66$ with only 2 Lorentzians). The typical frequency values are $\Delta f_1=0.26^{0.32}_{0.21}\ \rm Hz$ and $\Delta f_2=5.3^{9.5}_{3.1}\ \rm Hz$, and the total fractional variability is $18^{26}_{16}\%$. The fit values obtained in the 69-130 keV band are similar though the typical frequency found for $\Delta f_1=0.44^{0.76}_{0.26}\ \rm Hz$ is a bit higher and the overall variability decrease to $12^{14}_{10}\%$.

For the $SS_{ASM}$, only two Lorentzians are needed to fit well ($\chi^2/\nu=1.66$) the 27-49 keV PDS. The overall variability is at $11\pm 2\%$, hence lower than in $HS_{ASM}$, but the first Lorentzian frequency increased as $\Delta f_1=0.50^{0.94}_{0.27}\ \rm Hz$. In the 69-130 keV band, the variability remains at a level of $11\pm5\%$.
\subsubsection {States defined by SPI}  
Since the errors on $\Gamma_{ASM}$ are not negligible (see center panel of Fig. \ref{fig:rms-flux-hr-vs-time}) and make the previous observation grouping questionable, we also tried using the SPI hardness ratio for defining two different states:
\begin{itemize}
\item $HR_{SPI}>0.75$ corresponds to a hard state ($HS_{SPI}$), whereas,
\item $HR_{SPI}<0.75$ corresponds to a soft state ($SS_{SPI}$).
\end{itemize}
\begin{figure*}
\centering
\includegraphics[angle=-90,scale=0.4]{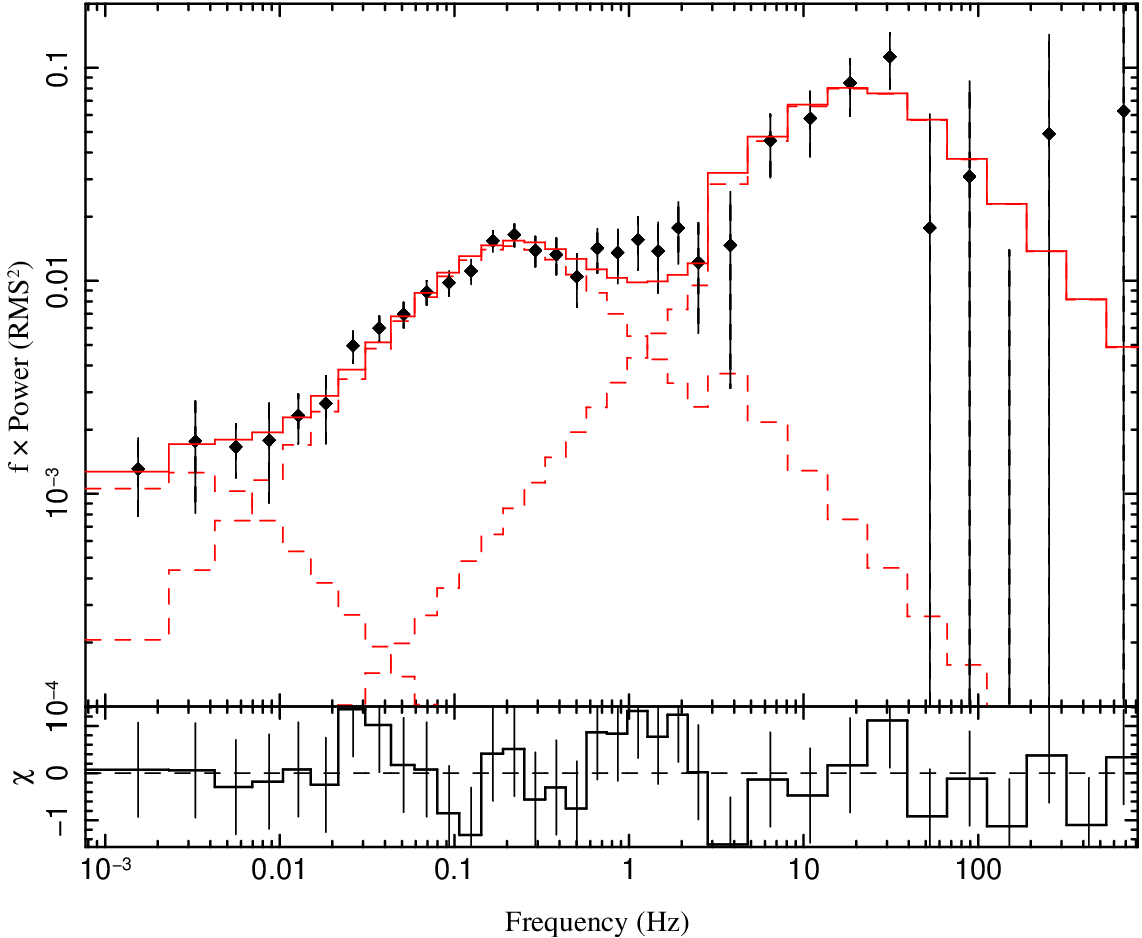}
\includegraphics[angle=-90,scale=0.4]{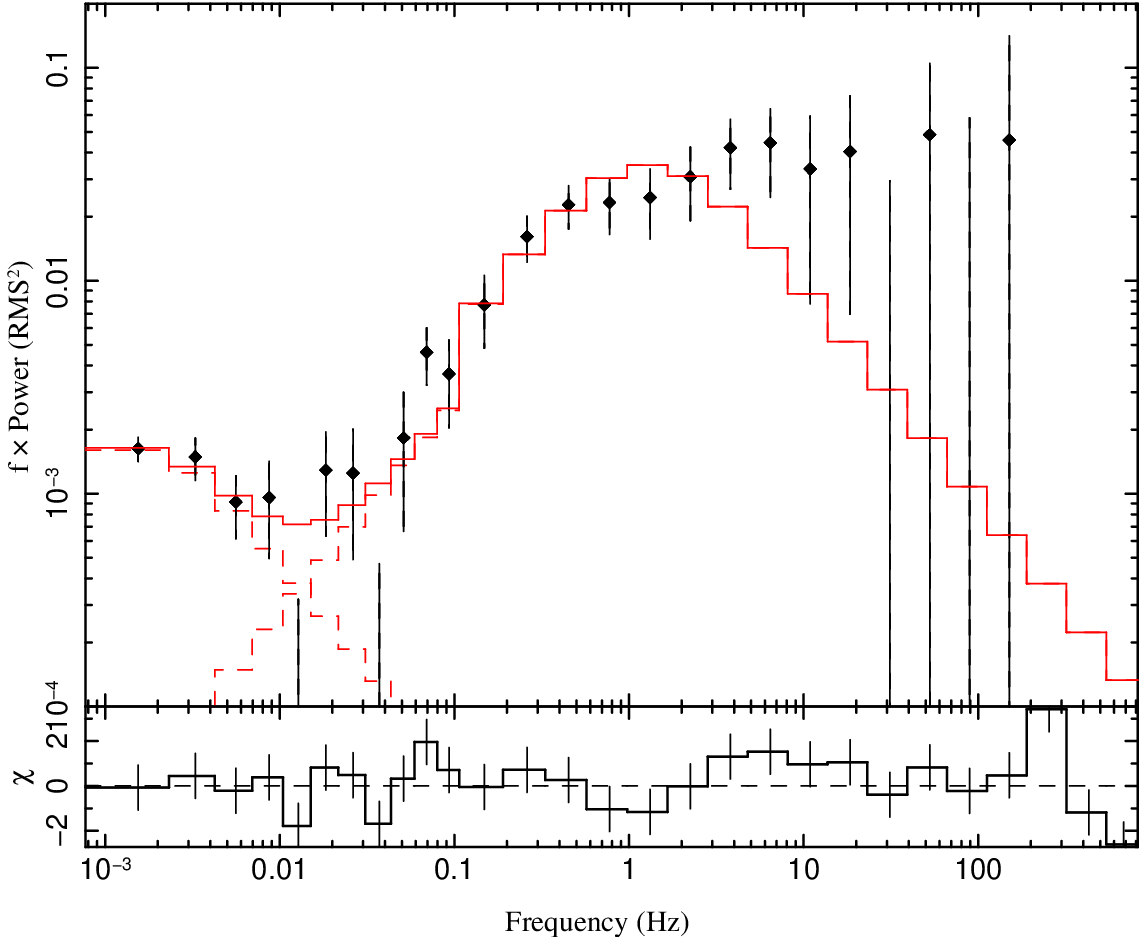}
\caption{Cyg X-1 PDSs calculated when the source was hard (left panel) and soft (right panel) according to SPI hardness ratio criterion in the 27-49 keV band.}\label{fig:psd-hard-soft}
\end{figure*}
This value of 0.75 was chosen because of the clear separation visible in $HR_{SPI}$ between two sets of data. Such a separation would correspond to a boundary value in ASM photon index of $1.80$ according to the previous $HR_{SPI}$ vs $\Gamma_{ASM}$ fit, but is compatible with the previous boundary of $\Gamma_{ASM}=2$, when considering the typical errors on this fit ($\Delta \Gamma_{HR_{SPI}=0} = 0.52$ at 90\%).  
 
The 27-49 keV band PDSs for both SPI defined states are plotted in Fig. \ref{fig:psd-hard-soft}. For the 27-49 keV PDSs, 3 Lorentzians were necessary to fit the $HS_{SPI}$ PDS, whereas only two where used for the $SS_{SPI}$ one.  The addition of a second component was tested for the $HS_{SPI}$ and the corresponding F-test NHP is equal to $6.3\ 10^{-7}$. The addition of a third one gives a F-test NHP equal to $1.5\ 10^{-3}$. For the $SS_{SPI}$, the addition of a second component (with parameter $\Delta f_1$) is highly required as it gives a F-test NHP equal to $2.7\ 10^{-6}$. The addition of a third Lorentzian in the $SS_{SPI}$ PDS would  is unlikely since the corresponding calculated F-test NHP is equal to $8\%$. For the 69-130 keV, only two Lorentzians were used for $HS_{SPI}$ and one for $SS_{SPI}$.

  \begin{deluxetable}{c|c||c c}
\tabletypesize{\scriptsize}
\tablecaption{SPI Hard state: results of the PDS fit by 3 Lorentzians. \label{tbl:hard-rms}}
\tablewidth{0pt}
\tablehead{
                & \colhead{27-49 keV} & \multicolumn{2}{c}{69-130 keV}
}
\startdata
                 &                   &  \tablenotemark{(a)}                              &  \tablenotemark{(b)}\\
          $\displaystyle \Delta f_0$ (Hz)    &  $\displaystyle  0.0028^{  0.30}_{ 0.0007  }$   &  $ \displaystyle   0.0030_{     1e-10     }^{    11.4 }$         & $ \displaystyle    0.0028\ (f) 			$           \\
        $\displaystyle rms_0$         &     $\displaystyle 0.040^{ 0.049}_{  0.028}$     &      $ \displaystyle  0.029_{ 0.013    }^{   0.029 } $          & $ \displaystyle    0.027  ^{   0.037	}_{    0.009 } 	$   \\
         $\displaystyle  \Delta f_1 $(Hz)    &    $\displaystyle  0.22^{  0.26}_{   0.18}  $    &     $ \displaystyle   0.40_{  0.23 }^{       0.70 }$            & $ \displaystyle     0.22\ (f) 			$           \\
       $\displaystyle  rms_1$         &    $\displaystyle  0.14^{  0.14}_{   0.13}$      &       $ \displaystyle  0.13_{ 0.10    }^{    0.15 }$           & $ \displaystyle    0.10  ^{    0.12  }_{    0.08 } 	$   \\
        $\displaystyle   \Delta f_2$ (Hz)    &    $\displaystyle 20.3^{31.8}_{12.3}$             & \nodata  	    & $ \displaystyle      20.3\ (f) 			$           \\
       $\displaystyle  rms_2$         &    $\displaystyle  0.22^{  0.26}_{0.19}$         &     \nodata            & $ \displaystyle     0.22  ^{     0.28	}_{     0.12 } 	$   \\
$ \displaystyle  rms_{tot}/mean$                           & $\displaystyle 0.26^{0.29}_{0.23}$                & $ \displaystyle 0.13_{0.10}^{0.15}$             &  $ \displaystyle  0.24^{0.30}_{0.16}$                 \\ 
$\displaystyle  \chi^2/\nu$           &      0.70                                         & 1.06                                            & 0.84                                  
\enddata
\tablecomments{$(f)$ stands for a parameter value fixed in the fit.}
\tablenotetext{(a)}{Values obtained when all parameters of the fit are let free to vary.}
\tablenotetext{(b)}{Values obtained when tying the  Lorentzian frequencies to the value obtained in the 27-49 keV band.}
\end{deluxetable}
 \begin{deluxetable}{c|c||c c}
\tabletypesize{\scriptsize}
\tablecaption{SPI Soft state: results of the PDS fit by 1 or 2 Lorentzians \label{tbl:soft-rms}}
\tablewidth{0pt}
\tablehead{
                & \colhead{27-49 keV} & \multicolumn{2}{c}{69-130 keV} 
}
\startdata
                  &                   &  \tablenotemark{(a)}                              &  \tablenotemark{(b)}\\
$\displaystyle  \Delta f_0 $(Hz)             & $ \displaystyle       0.0015  ^{    0.0023	 }_{  0.0008} $  & \nodata			    & $ \displaystyle  0.0015\ (f)$ 			          \\ 
$\displaystyle  rms_0$                & $ \displaystyle         0.044  ^{     0.049	}_{     0.040} $  & \nodata			    & $ \displaystyle  0.043 ^{    0.062    }_{         0} $      \\   
$\displaystyle \Delta f_1$ (Hz)              & $ \displaystyle          1.28  ^{       2.27	  }_{    0.77} $  & $ \displaystyle    0.26  ^{     1.86 }_{ 0.11} $ & $ \displaystyle  1.28\ (f)$ 			          \\
$\displaystyle rms_1$                 & $ \displaystyle         0.15  ^{      0.17	 }_{     0.13} $  & $ \displaystyle   0.20  ^{    0.39 }_{  0.14}  $ & $ \displaystyle   0.29 ^{     0.34  }_{    0.22} $         \\  
$ \displaystyle  rms_{tot}/mean$                           & $\displaystyle 0.15^{0.18}_{0.13}$                & $ \displaystyle   0.20_{0.14}^{0.39}$           & $ \displaystyle   0.29   ^{    0.34  }_{   0.23}$           \\
$\displaystyle  \chi^2/\nu$           &      1.74                                         &   0.99                                          &   0.96                                
\enddata
\end{deluxetable}

 The results of the fits of both states, are given in Tables \ref{tbl:hard-rms} and \ref{tbl:soft-rms}. It shows first that the $HS_{SPI}$ overall variability is quite high in the 27-49 keV band ($26\pm 3\%$), and dominated by a high frequency ($\sim 20\rm \ Hz$) component. As this latter component is not needed in the 69-130 keV PDS, the variability drops down to $13\pm 2\%$. Tying the frequency values in the 69-130 keV PDS to the one obtained in the 27-49 keV band lead however to a rms value of $24^{30}_{16}\%$ consistent with the one obtained at low energy.

In the $SS_{SPI}$, the frequency of the first Lorentzian increases up to $\Delta f_{1}=1.3\ \rm Hz$. The variability is lower than in $HS_{SPI}$ ($\sim 15\%$) in the 27-49 keV band, but it increases in the 69-130 keV band (up to $\sim20\%$ or even $\sim30\%$ when the frequencies are tied to the ones obtained at low energy). 

Comparing to the previous state grouping based on ASM criterion, it is interesting to note that the typical frequencies obtained here are more different between the hard and the soft state. For example the frequency $\Delta f_1$ in $SS_{ASM}$ is only twice the one in $HS_{ASM}$, whereas there is a factor 7 for $\Delta f_1$ between $SS_{SPI}$ and  $HS_{SPI}$. A difference in the values of $\Delta f_2$ is also noted between $HS_{SPI}$ (where $\Delta f_2\sim 20\ \rm Hz$ is high) and $HS_{ASM}$ (where $\Delta f_2\sim 5\ \rm Hz$ is low).
\subsection{Comparison with {\it RXTE}/PCA power spectra}
We also compared our {\it INTEGRAL}/SPI PDSs with {\it RXTE}/PCA ones. The only contemporary {\it RXTE} observations found for our dataset spans from revolution \# 628 to 630 for {\it INTEGRAL}. During those observations, the average ASM photon index is hard as $\Gamma_{ASM} = 1.8\pm0.15$. We retrieved and analyzed in turn the whole observation \#93121 for {\it RXTE}/PCA. The PCA observation is simultaneous to {\it INTEGRAL} ones, but the duration of the latter is larger for increasing the statistics. For the PCA observations, event files were processed in the standard way with {\verb seextrct } and background lightcurves were calculated with {\verb pcabackest } and {\verb saextrct }. We finally computed the PCA PDS using {\verb powspec } on the 20-40 keV background substracted lightcurves.
  
\begin{figure*}
\centering
\includegraphics[angle=-90,scale=0.4]{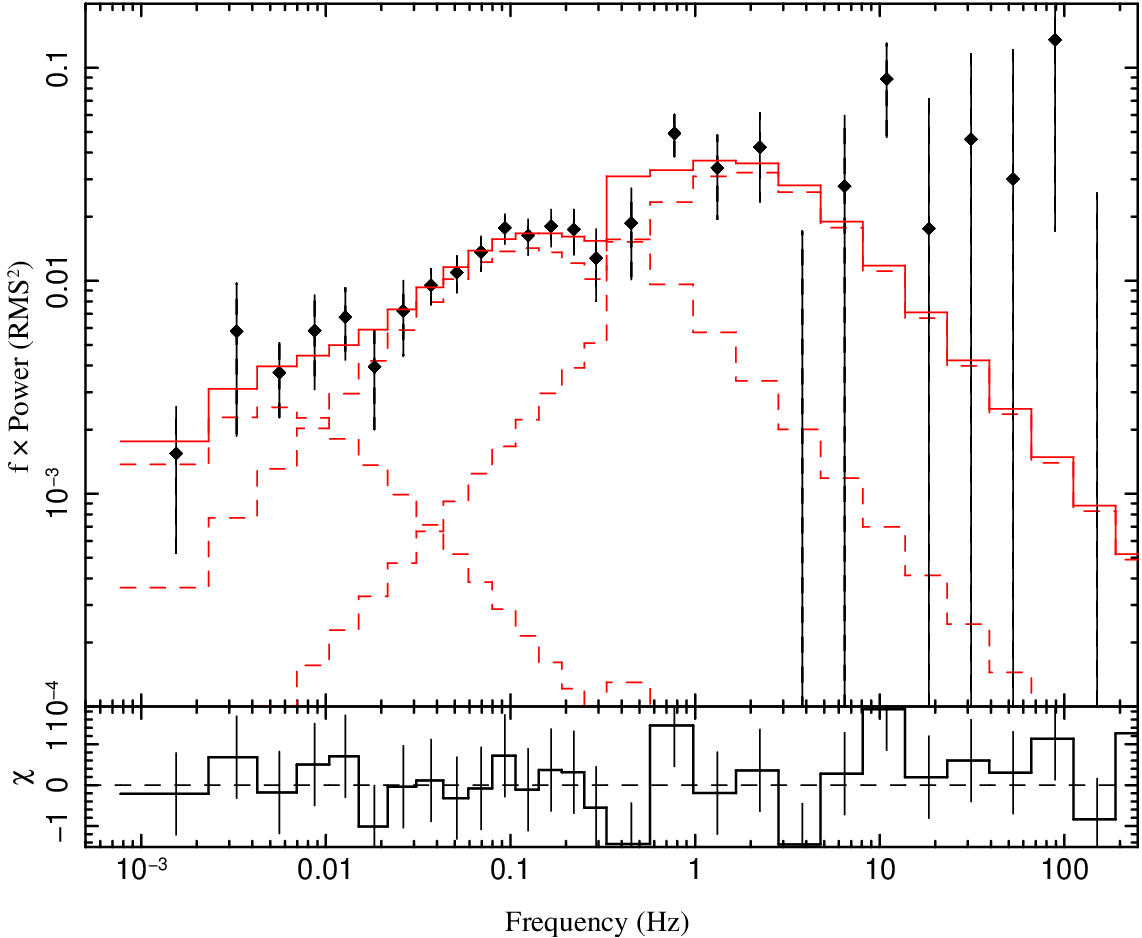}
\includegraphics[angle=-90,scale=0.4]{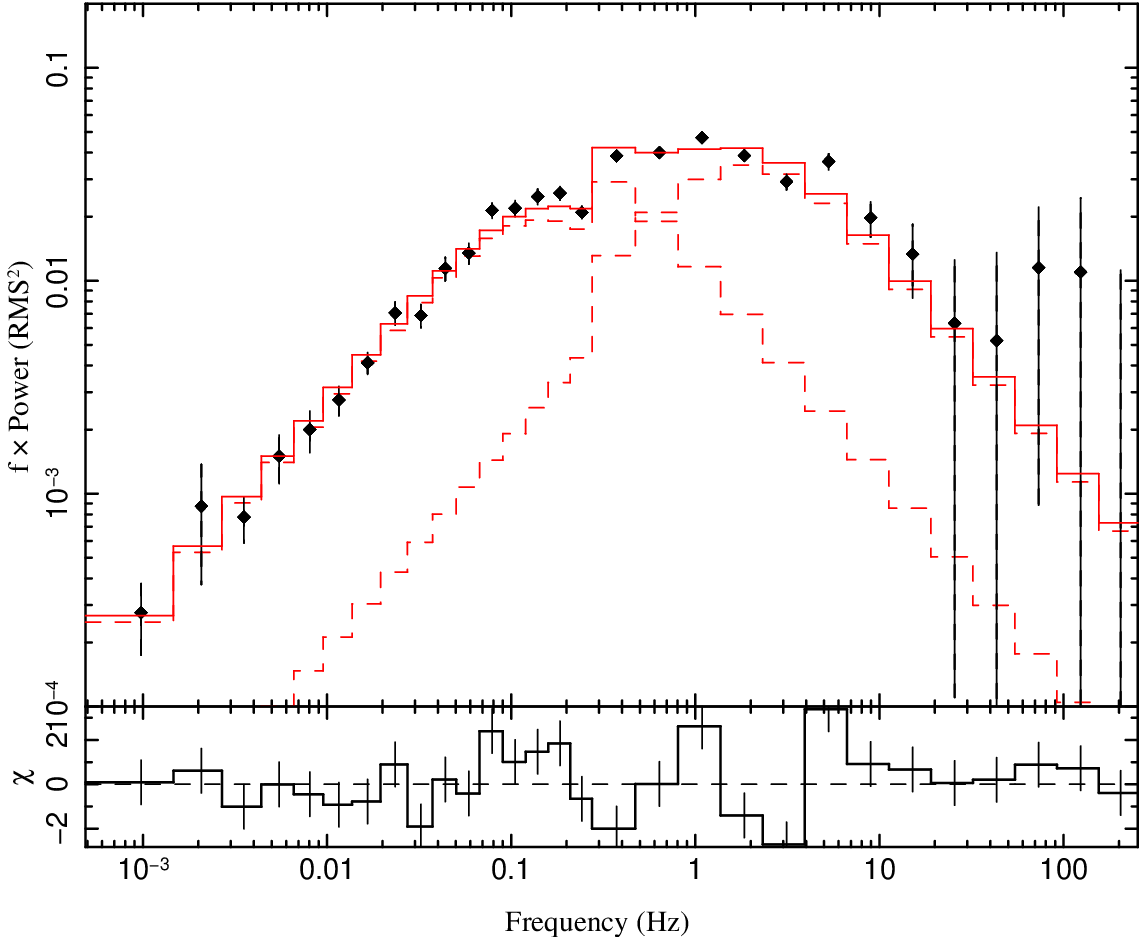}
\caption{Cyg X-1 PDS for SPI in the 27-49 keV band (left panel) as compared to contemporary PDS in the 20-40 keV with the PCA  (right panel). See text for the result of the fit.}\label{fig:psd-SPI-PCA}
\end{figure*}
Both PDSs are plotted on Fig. \ref{fig:psd-SPI-PCA}. It is first clear that the overall shape is similar from one instrument to the other. When fitted with two Lorentzians, the typical frequencies are indeed compatible, together with their normalization (see Table \ref{tbl:psd-SPI-PCA}). The only noticeable difference is the presence of a very low frequency complex present in SPI PDS whereas it is absent in the PCA ones. This absence could reflect the shorter duration of the PCA observation, and there would be not enough variability at frequencies comparable to its duration. In any case, setting the SPI PDS fit parameters to the value obtained for the PCA PDS still gives an acceptable fit as in that case $\chi^2/\nu=1.35$. 

\begin{figure}
\centering
\includegraphics[scale=0.45]{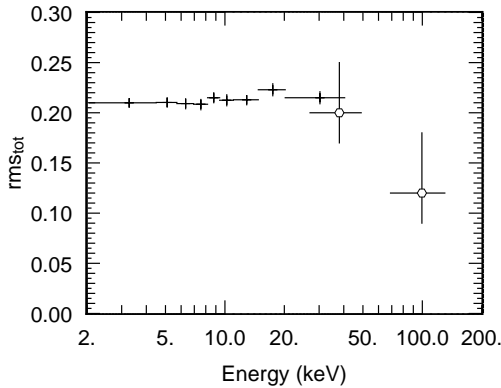}
\caption{Cyg X-1 rms spectrum for the contemporary {\it INTEGRAL}/{\it RXTE} observations. Crosses = PCA, diamond = SPI data. 90\% errors plotted.}\label{fig:rms_spectrum}
\end{figure}

\begin{deluxetable}{c|c||c}
\tabletypesize{\scriptsize}
\tablecaption{Comparison between SPI and PCA PDSs: results of the fit \label{tbl:psd-SPI-PCA}}
\tablewidth{0pt}
\tablehead{
                & \colhead{SPI 27-49 keV} & \colhead{PCA 20-40 keV} 
}
\startdata
$\displaystyle  \Delta f_0 $(Hz)             & $ \displaystyle        0.005 ^{    0.100	 }_{ 0.001 } $    & \nodata 			          \\ 
$\displaystyle  rms_0$                & $ \displaystyle         0.06  ^{     0.07	}_{     0.03} $      & \nodata   \\   
$\displaystyle  \Delta f_1 $(Hz)             & $ \displaystyle        0.12 ^{    0.17	 }_{ 0.08 } $    & $ \displaystyle  0.151^{0.164}_{0.139}$ 			          \\ 
$\displaystyle  rms_1$                & $ \displaystyle         0.13  ^{     0.14	}_{     0.10} $      & $ \displaystyle  0.157 ^{    0.162    }_{ 0.151        0} $      \\   
$\displaystyle \Delta f_2$ (Hz)              & $ \displaystyle          1.8  ^{   16.7	  }_{    0.8 } $  & $ \displaystyle    1.9  ^{     2.2 }_{ 1.6} $  \\
$\displaystyle rms_2$                 & $ \displaystyle         0.14  ^{      0.21	 }_{     0.10} $  & $ \displaystyle   0.147  ^{    0.153 }_{  0.142}  $         \\  
$ \displaystyle  rms_{tot}/mean$                           & $\displaystyle 0.20^{0.25}_{0.17}$                          & $ \displaystyle   0.215   ^{    0.221  }_{   0.209}$           \\
$\displaystyle  \chi^2/\nu$           &      0.75                                                                                   &    1.87         
\enddata
\end{deluxetable}
Finally, for the 69-130 keV band, due the lower statistics, only one Lorentzian is required for fitting the PDS ($\chi^2/\nu=1.04$). The typical frequency is consistent with the ones obtained at lower energy since $\Delta f_{1,\ 69-130\ keV}=0.19^{0.46}_{0.08}$, and its variability is $rms_{1,\ 69-130\ keV}=0.12^{0.18}_{0.09}$. As the lower energy PCA PDSs were also analysed for this observation, for the first time a 2-130 keV high frequency rms spectrum of the source was build. It is plotted in Fig. \ref{fig:rms_spectrum}. Note that $HR_{SPI}=0.84\pm0.01$ for this observation, hence Cyg X-1 was in a hard state at this time. The rms spectrum is flat until 49 keV ($\sim 20\%$), and decreases in the 69-130 keV bin.
Fitting the rms spectrum with a nearly flat powerlaw of index $\alpha=0.015^{0.029}_{0.002}$ gives a bad $\chi^2/\nu=2.4$, and the addition of a high energy cutoff improves the fit as $\chi^2/\nu=0.95$. The ftest NHP for such addition is however $1.5\%$, hence its presence is rather marginal, together with the lack of constraints on the fitted folded $E_{fold}=31^{122}_{20}\ \rm keV$ and cutoff $E_{cut}=84^{198}_{43}\ \rm keV$ energy values. The latter statistic is also dominated by the last bin (69-130 keV rms), as the 2-49 keV rms spectrum is well fit by a simple powerlaw ($\chi^2/\nu=0.94$).
\subsection{Time lags}
\begin{figure*}[t]
\centering
\includegraphics[angle=-90,scale=0.3]{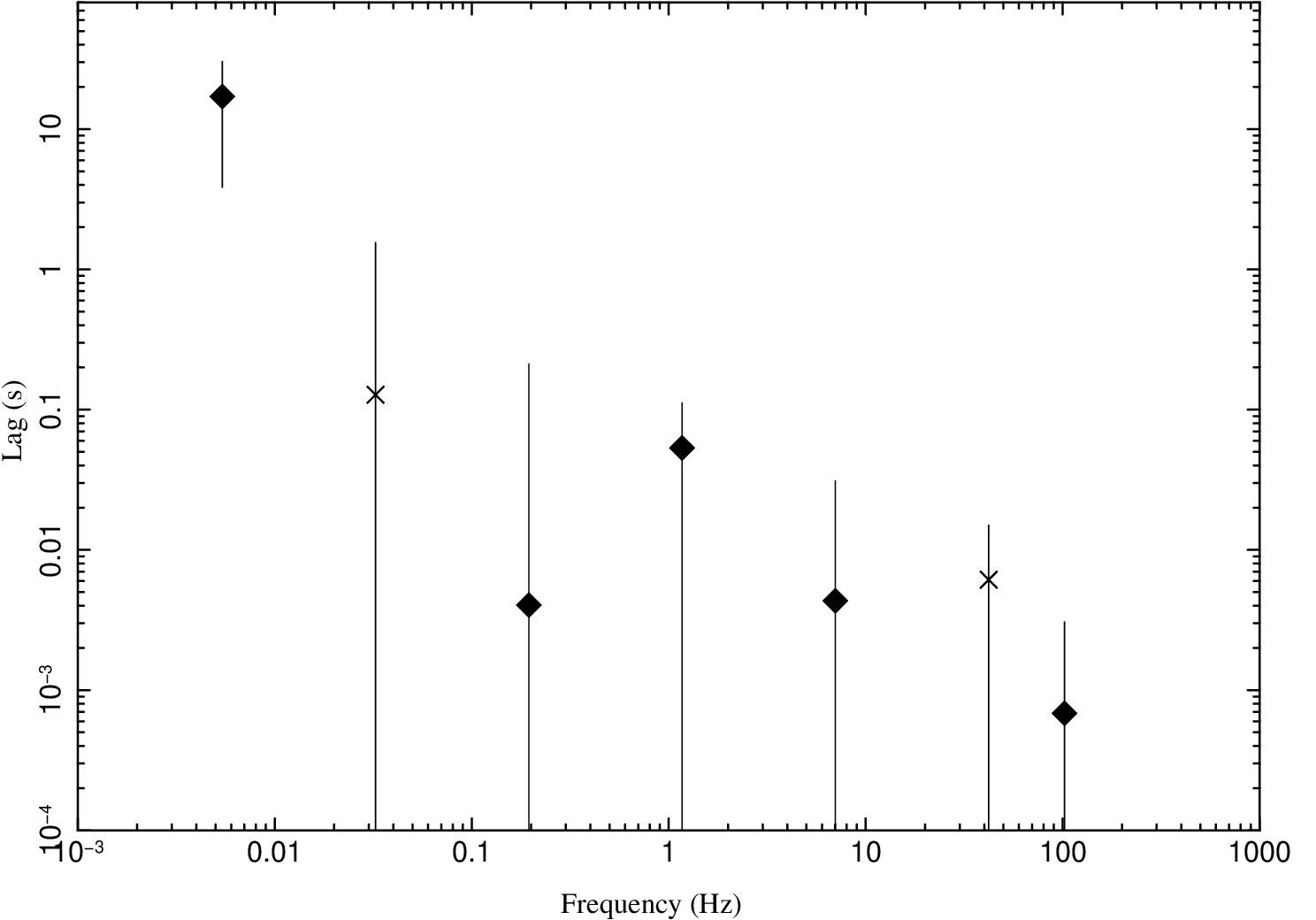}
\includegraphics[angle=-90,scale=0.3]{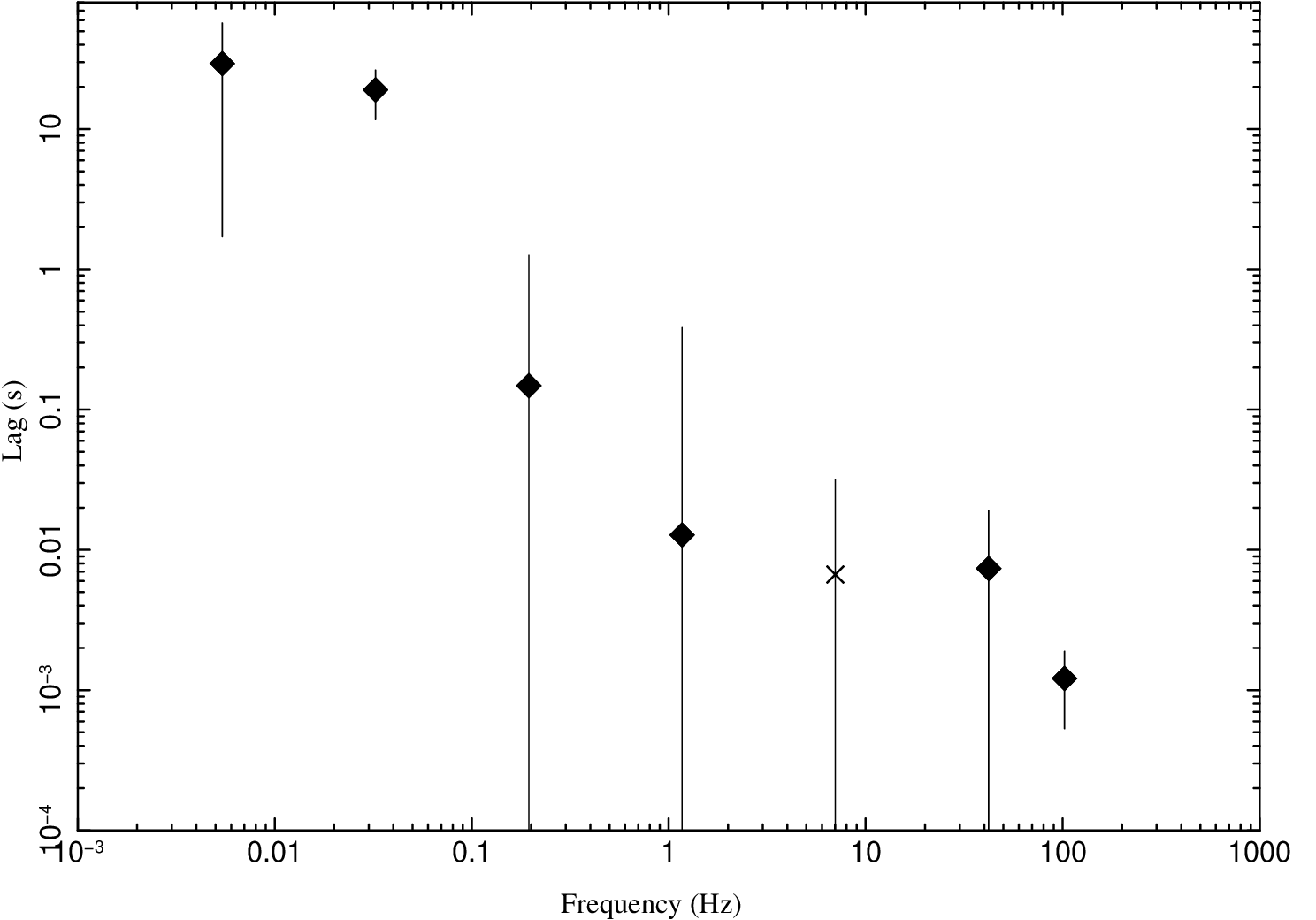}
\caption{Lags between the 69-130 keV and the 27-49 keV band as a function of energy for the hard (left panel) and soft state (right panel). Filled diamonds stand for 69-130 keV band lagging the 27-49 keV bands, i.e., positive lags, whereas crosses stands for negative ones.}\label{fig:lags-hard}
\end{figure*}

We looked for time lags between the low energy band (27-49 keV) and the hard one (69-130 keV). For that purpose, we used the {\verb sitar_lags }\ function provided by the {\verb SITAR }\ package. Lags as a function of frequency were then obtained and are plotted in Figs. \ref{fig:lags-hard}. It appears first that the obtained lags are dominated by the Poisson noise due to the background contribution. 

 In the hard state, a strong rebinning of the cross-power spectrum tends to  show that the calculated lags are consistent with 0 or slightly positive (i.e., 69-130 keV band lags the 27-49 keV band) at low frequency (i.e $<10^{-2}\ \rm Hz$). The effective lag calculated is e.g., $\Delta \tau = +18\pm 12\ \rm s$ ($1 \sigma$ errors) in the   $[1.5,9.2] 10^{-3}\ \rm Hz$ frequency band. Between 1 and 10 Hz, an upper limit of $\Delta \tau < +30 \rm \ ms$ can be set. This limit is compatible with the positive lags observed in softer X-ray in the hard state, where it is typically $\Delta \tau \sim 1-4\rm\ ms$ (see e.g. \citealp{pottschmidt2003}). When integrated over a wide frequency range, cross-power spectra are consistent with 0 lags at (90\%) for this hard state: we indeed obtain  $\Delta \tau = +6.8\pm 6.4 \rm\ s$ ($1 \sigma$ errors) in the $[00.0015,1.5]\ \rm Hz$ frequency range.
In our soft state, the calculated lags seem also positive and slightly better constrained at low frequencies.  For example, $\Delta \tau=+19\pm7\ \rm s$ in the $[0.92,5.5]\ 10^{-2}\ \rm Hz$ frequency band. As for the hard state, an even stronger rebinning of the cross-power spectrum results in obtaining a value consistent with 0 lags (at 90\%) as in that case $\Delta \tau = 35\pm 29\rm s$ ($1 \sigma$ errors) in the $[00.0015,1.5]\ \rm Hz$ frequency range. Those results are consistent with those obtained by \cite{torii2011}, as their cross-correlation functions between the 10-60 keV and 60-200 keV peak at 0 lag.

\section{Discussion \& Conclusions}
SPI on-board {\it INTEGRAL} was used successfully to study Cygnus X-1 variability at high energies. The high stability of the instrument allows to probe Cyg X-1 variability up to 100 Hz and down to 1 mHz. The first and main result of this study concerns the evolution of the variability at high energies: it is indeed demonstrated that Cyg X-1 is varying at a high level ($\sim 25\%$) up to 130 keV.

 The shapes of the high energy PDS obtained are similar to that obtained at lower energies with other instruments as it is well described by subset of up to 4 Lorentzians. High frequency variability ($\sim 40 \rm Hz$) is observed in the 69-130 state averaged PDSs, showing for the first time that fast variability is present at high energy. 
 

The parameters of the Lorentzians used are also evolving with the source state. For example, in order to explain the differences between the $HS_{SPI}$ and $SS_{SPI}$ frequencies found, the simplest solution is to associate those frequencies with given physical components, at certain typical radii in the disk or in the corona. In our case, the first Lorentzian is shifted from $1.6\rm Hz$ in the $SS_{SPI}$ to $0.22\rm Hz$ in the $HS_{SPI}$. If this component is linked to Keplerian motion, it would imply that the typical radius for this component increases by a factor $(0.22/1.6)^{-2/3}=3.8$ between the soft and hard state. In the framework where the variability is driven by waves propagating in the corona (see. e.g. \citealp{cabanac2010}), the typical break frequency scales as the inverse of the external corona radius. Hence in that latter framework the corona external radius would be 7 times larger in our $HS_{SPI}$ than in the $SS_{SPI}$.
It is interesting to note that the frequency shifts for $f_1$ follow the  difference between the hard and soft state fundamental planes found by \cite{koerding2007}: this latter is based on the fundamental plane found by \cite{mchardy2006} which links, for SS BHB and AGN, the mass accretion rate $\dot{m}$, the first Lorentzian frequency $f_1$ and the black hole mass $M$ with $f_1 \times M \propto \dot{m}$. Moreover, \cite{koerding2007} have shown that this relation still holds in the HS, at price of changing the proportionality factor.  Based on an extensive analysis, they calculated that this factor was equal to 0.3 for HS BHB, whereas it was 2.2 in the SS. In our case, we note that the ratio between $\Delta f_{1,\ soft}$ and $\Delta f_{1,\ hard}$ is equal to 7, which is therefore very close to the proportionality factor ratio in both fundamental planes. 
 Such different tracks for HS and SS states in the fundamental plane is interpreted by \citeauthor{koerding2007} as a sign that not only the accretion rate is driving the accretion process evolution. An additional parameter must be invoked to render this two different tracks in the Hard and Soft states. 

The rms evolution with the hardness of the source that we observed at high energy is similar to the rms-hardness correlation seen in many XRB (see e.g., the Hardness-RMS Diagrams in \citealp{belloni2010}). It reflects the known relation that the source gets more variable when it gets harder (see however some high variable soft state in GZD10, Fig. 2b). Our results support the fact that this relation still holds at high energies, giving strict constrains on emission models that intend to reproduce the observed variability.

The rms level we obtained ($\sim 25\%$) together with the frequency of the first Lorentzian ($\Delta f_1\sim0.2\ \rm Hz$) in our hard state  is compatible with the majority of the hard states found at lower energies with the PCA in GZD10 (see e.g. Figs 2b and 7a of that paper). If in GZD10 the hardest states rms reaches 35-40\%, our hard state level, PDS and typical frequencies are compatible with bit softer ones. In particular, our hard state PDS (Fig.  \ref{fig:psd-hard-soft}) is close in shape to the PCA PDS for observation 60090-01-19-00 (Fig 6b of GZD10), for which a rms spectrum is computed: the variability is rather constant around 25-30\% rms on the 2-30 keV band, and adding our two high energy point would show that the rms remains at roughly the same level in the 2-130 keV band for this hard state.
This rather constant rms spectrum in the hard state is emphasized by our simultaneous PCA/SPI analyse of the source, where the source was hard. If a break in the hard X-ray ($\sim 80\ \rm keV$) is possible in this rms spectrum, its significance is rather marginal. It is also possible that the low duration of this observation and hence the lack of statistics does not allow to constrain a putative high frequency tail, which could account for extra variability. Independent results from \cite{torii2011} obtained with {\it Suzaku} showed that the hard state of Cyg X-1 rms is indeed rather constant in the 12-200 keV energy range.
 
Our $SS_{SPI}$ can be similarly associated with PCA observation id 70104-01-03-00 analysed by GZD10. The rms spectrum is in turn also quite constant between 2-35 keV, though its level is lower than in the previous hard state. Our SPI level of variability in the 27-49 keV band is consistent with the one obtained in the highest energy band of this PCA observation (15-20\%) and it increases in the 69-130 keV (20-30\%).
An interesting interpretation of the rms spectra has been attempted by \cite{gierlinski2005} and more recently in GZD10, with the following principles: any lightcurve variability measured by its fractional rms in a given energy band may be due to the variability of at least one of the parameters $p_i$ of the average spectrum. \cite{gierlinski2005} and GZD10  used the combination of a multicolor disk ({\verb diskbb }\ \citealp{mitsuda1984}), the {\verb eqpair }\  hybrid comptonization model \citep{coppi1999} and a reflection component \citep{magdziarz1995}, in order to model the average spectrum $F$. The effect on the output rms spectrum of each parameter variation is then investigated. Qualitatively, if a parameter $p_i$ only impacts on the normalization of a model, the spectrum will be shifted by a certain factor, independently from the energy. Hence, the variability of this parameter will lead to a flat rms spectrum. On the contrary, if another parameter $p_j$ (e.g. the photon index $\Gamma$) changes the slope of the spectrum around a pivot, the factor by which the spectrum is shifted will not be energy independent: the further the working energy band is from the pivot, the higher the rms.

For the hard state, a flat rms spectrum is observed for XTE J1550-564 and XTE J1650-500 \citep{gierlinski2005} and Cyg X-1 (GZD10). In those papers, it is proposed that the flat rms spectrum in the hard state may be reproduced by the number of soft photon variability or fluctuation of the accretion rate. Given that our rms spectrum behaviour remains flat until 130 keV in the hard state, we demonstrate that this interpretation still holds at high energies.

For our soft state examined here, the rms spectrum which is flat until 50 keV, may then increase at high energies. Following \citep{gierlinski2005}, variability of the injection Lorentz factor $\Gamma_{inj}$ may reproduce such rms spectrum increase at high energies, but a maximum should also be observed around 5-10 keV.

The above explanation of rms spectra however first depends on the underlying average model chosen, and second on the fact that only one parameter variation is considered. It is hence possible that some of the parameters are linked by underlying physical processes, such as e.g. the luminosity of the hard and soft components when $\dot{m}$ is varying. As an example, \cite{cabanac2010}, in the framework of an oscillating corona model, examined the impact of an adiabatic oscillation on the rms spectra. Both the corona temperature and the optical depth are then varying, which in turns leads to a Compton $y$ parameter variation. They predicted the presence of a minimum in the rms spectrum, due to the pivoting around the average spectrum. The position of the pivot depends on the relative peak between the soft photon source and the comptonized component.
     
We finally note that the high energy band power spectra that we obtained exhibits peaked-noise components, one of wich (the lower frequency one) is consistent with the QPO detected with SIGMA by \cite{vikhlinin1994} in terms of peak frequency value. Those components are sharper in the higher than the lower energy PDSs (comparing e.g. Figs. \ref{fig:psd-301-684} and \ref{fig:psd-301-684-69-130}). This effect is also observed at lower energies with {\it RXTE} (see e.g. Fig. 2 of \citealp{cui1997}). If confirmed, it may therefore indicate that the Lorentzians  quality factor increases with energy. In other words, the coherence of the underlying variability processes would increase with energy, giving constraint on the emission models.
\acknowledgements
The {\it INTEGRAL} SPI project has been completed under the responsibility and leadership of CNES. We are grateful to ASI, CEQ, CNES, DLR, ESA, NASA and OSTC for support.

CC acknowledges M. Del Santo for having kindly provided IBIS Cyg X-1 field of view during EXO 2030+375 giant flare, and R. Belmont for stimulating discussions. The authors thank the anonymous referee for improving the quality of the paper. CC gratefully acknowledges financial support provided by CNES.  
\bibliographystyle{apj}
\bibliography{art_ref}
\appendix
\section*{Contribution of other sources in the FOV : deconvolution algorithm for PDS with SPIROS}\label{sect:deconvolution}
Our deconvolution algorithm is based on the fact that the Leahy Power is a linear function of the count rate. As a consequence, any deconvolution algorithm can be used for the discrimination between the sources PDSs, if the PDSs are Leahy renormalized. Methods using either PIF (Partial Illumination Fraction of a detector) or usual deconvolution algorithm can be used. PIF methods (see e.g. an example applied to IBIS/ISGRI data, \citealp{cabanac2006}) are quite relevant when the detector number of pixels is large. For SPI, since the pixel number is low (19), usual deconvolution algorithm was preferred here. Hence, first for each SCW, 19 different PDSs are calculated (after retrieving and selecting the events according to the energy band chosen).
Second SPIROS (ISDC standard deconvolution algorithm \citealp{skinner2003}) is used in spectral mode, but the input files are modified: counts per energy band and detector are usually employed, but they are here replaced by the Leahy Power per frequency bin and detector. The same process was applied to the background files. We can therefore summarize this process by the following substitution in the input files:
\begin{eqnarray}
Energy\ bin&\longrightarrow& Frequency\ bin\\
Detector\ Counts&\longrightarrow& Leahy\ Power\times Livetime\\
Back.\ Counts&\longrightarrow& Noise\ Leahy\ Power\times Livetime.
\end{eqnarray}
The multiplication of the Leahy Power by the actual live-time of the observation is due to the fact that the Leahy Power scales as the count rate, whereas the usual deconvolution algorithm uses the total counts.
 We finally ran SPIROS in background handling mode 3, and obtained the deconvolved PDSs for for Cyg X-1 and EXO 2030+375.
\begin{figure}[t]
\centering
\includegraphics[angle=-90,scale=0.4]{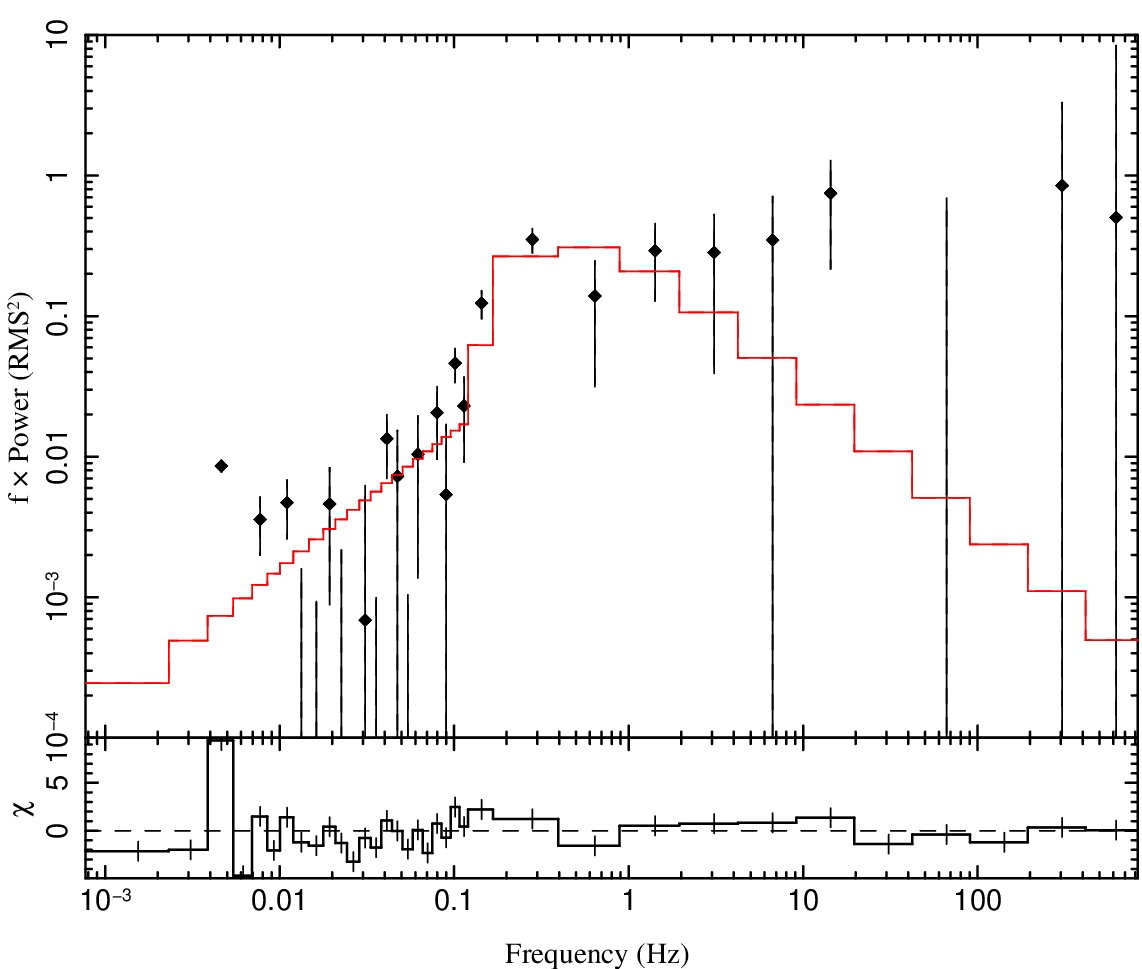}
\includegraphics[angle=-90,scale=0.4]{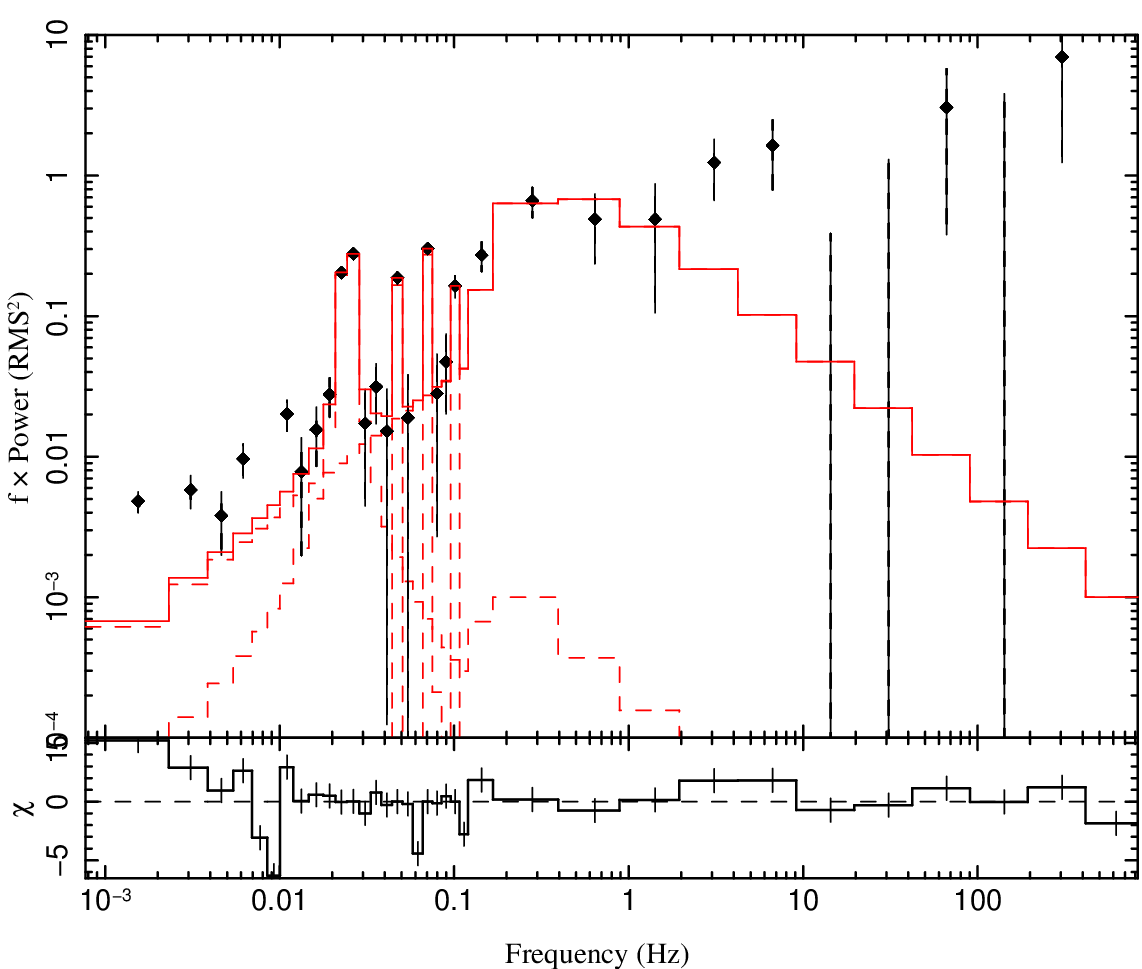}
\caption{Deconvolved (and noise subtracted) rms normalized PDSs of Cyg X-1 ({\it up}) and EXO 2030+375 (bottom) calculated from rev \#452 to 486 in the 27-49 keV band. Note the pulse fundamental and its harmonics well associated with EXO 2030+375, and not present in Cyg X-1 PDS as expected.}\label{fig:psd-deconv}
\end{figure}
They are plotted in Fig. \ref{fig:psd-deconv}. In our sample, the EXO 2030+375 pulse and its 2 first harmonics are clearly detected in the associated PDS whereas it is absent in Cyg X-1 PDS as expected. A fit of the pulse and its harmonics by narrow Lorentzians gives a peak frequency of $f_{pulse}=24 \rm\ mHz$, which corresponds to a pulse period of $T_{pulse}=41.6 \rm s$. This value is consistent with the one derived by other instruments/observatories for the 2006 outburst (see e.g. \citealp{klochkov2007}). None of these features are present in the Cyg X-1 PDS, which gives good confidence in our deconvolution algorithm.
 
We also fitted the continuum with one zero centered Lorentzian. During these observations, we estimated the fractional variability of Cyg X-1 to be $rms_1=rms_{tot}/mean=25^{30}_{21}\%$. For EXO 2030+375, fitting the PDS continuum with a single zero centered Lorentzian gives a rather bad fit ($\chi^2/\nu=1.91$), but gives an evaluation of the low frequency variability with  $rms_1=35^{39}_{31}\%$. Finally, the Leahy level obtained for EXO 2030+375 during its giant outburst is comparable to the one obtained for Cyg X-1. Hence, given that EXO 2030+375 flux is usually $\sim 10$ times fainter\footnote{see e.g. the {\it Swift}/BAT lightcurve on \url{http://heasarc.gsfc.nasa.gov/docs/swift/results/transients/EXO2030p375/.}}, we confirmed that our first approximation of neglecting this source in the raw PDS calculation is valid. The rms value calculated for Cyg X-1 in the previous sections will then be correct within $\sim 3\%$ systematic uncertainties.  
\end{document}